\documentclass[final,3p]{elsarticle}

\usepackage{amscd,amsmath,amsthm,amsfonts,amssymb}
\usepackage{graphicx}
\usepackage{multicol}
\usepackage{color}

\numberwithin{equation}{section}

\newcommand{\comment}[1]{}
\newcommand{\be}{\begin{equation}}
\newcommand{\ee}{\end{equation}}

\newtheorem{theorem}{\bf Theorem}[section]

\def\tanh{{\rm tanh}}

\journal{Physica D}
\begin{document}
\begin{frontmatter}
\title{Collapse for the higher-order nonlinear Schr{\"o}dinger equation}
\author[ath]{V. Achilleos}

\author[aegean]{S. Diamantidis}

\author[ath]{D. J. Frantzeskakis}

\author[ioa]{T. P. Horikis\corref{cor1}}

\author[aegean]{N. I. Karachalios}

\author[umass]{P. G. Kevrekidis}

\address[ath]{Department of Physics, University of Athens, Panepistimiopolis, Zografos, Athens 15784, Greece}

\address[aegean]{Department of Mathematics, University of the Aegean, Karlovassi, 83200 Samos, Greece}

\address[ioa]{Department of Mathematics, University of Ioannina, Ioannina 45110, Greece}

\address[umass]{Department of Mathematics and Statistics, University of Massachusetts, Amherst MA 01003-4515, USA}
\cortext[cor1]{Corresponding author}
\begin{abstract}
We examine  conditions for finite-time collapse of the solutions of
the higher-order nonlinear Schr\"{o}dinger (NLS) equation incorporating
third-order dispersion, self-steepening, linear and nonlinear gain
and loss, and Raman scattering; this is a system that
appears in many physical contexts as
a more realistic generalization of the integrable NLS. By using
energy arguments, it is found that the collapse dynamics is chiefly controlled by the linear/nonlinear gain/loss
strengths. We identify a critical value of the linear gain, separating the possible decay of solutions to the trivial zero-state,
from collapse. The numerical simulations, performed for a wide class of initial data, are found to be in very good agreement
with the analytical results, and reveal long-time stability properties of localized solutions. The role of 
the higher-order effects to the transient dynamics is also revealed in these simulations.
\end{abstract}

\begin{keyword}
Collapse \sep instabilities \sep solitons \sep nonlinear optics
\PACS 42.65.Sf \sep 42.65.Re \sep 02.30.Jr
\end{keyword}

\end{frontmatter}

\section{Introduction}
The present paper deals with the dynamical properties of a higher-order nonlinear Schr\"{o}dinger (NLS) equation,
expressed in dimensionless form:
\begin{eqnarray}
\label{eq1}
{{u}_{t}}+\frac{s}{2}\mathrm{i} {{u}_{xx}}-\mathrm{i}|u|^2u=\gamma u+\delta|u|^2u+\beta u_{xxx}
+\mu(|u|^2u)_x+\left(\nu-\mathrm{i}\sigma_R\right)u(|u|^2)_x,
\end{eqnarray}
for all $x\in\mathbb{R}$, and $t\in [0,T]$ for some $T>0$. In Eq.~(\ref{eq1}), $u(x,t)$ is the unknown complex field, subscripts denote partial derivatives,
$\gamma$, $\delta$, $\sigma_R$ and $\mu,\,\nu$ are real constants, while $s = 1$ corresponds to
normal group velocity dispersion. We supplement Eq.~(\ref{eq1}) with the periodic
boundary conditions for $u$ and its spatial derivatives up to the second order:
\begin{equation}\label{eq2}	
u(x+2L,t)=u(x,t),\;\;\mbox{and}\;\;\;
\frac{\partial^ju}{\partial x^j}(x+2L,t)=\frac{\partial^ju}{\partial x^j}(x,t),\;\; j=1,2,\
\quad\forall\,x\in{\mathbb R},\;\;\forall\,t\in [0,T],
\end{equation}
where $L>0$ is given, and with the initial condition
\begin{equation}\label{eq3}
u (x,0)=u_0(x),\quad \forall\,x\in{\mathbb R},
\end{equation}
also satisfying the periodicity conditions~(\ref{eq2}). 
Note that the boundary conditions (\ref{eq2}) will be incorporated in the Sobolev spaces of periodic functions, that will be used for the analysis of the problem (\ref{eq1})-(\ref{eq3}). 
Their definition will be recalled and discussed below.

Equation (\ref{eq1}) finds
important applications in distinct physical and mathematical contexts,
with the most prominent being that
of nonlinear optics. Indeed, such models have been used to describe ultrashort
pulse propagation in nonlinear optical fibers in the framework of
the slowly-varying envelope approximation \cite{KodHas87,Agra1,Agra2} (see also reviews~\cite{Mih1,Mih2} and
references therein for recent work in few-cycle pulse propagation).
In such a case, $t$ and $x$ in Eq.~(\ref{eq1}) play,
respectively, the role of the propagation distance along the optical fiber and the retarded time (in a reference frame moving
with the group velocity), while $u(x,t)$ accounts for the (complex) electric field envelope.
The propagation of 
optical pulses (of temporal width
comparable to the wavelength) is accompanied by effects, such as
higher-order (and, in particular, third-order) linear dispersion (characterized by the coefficient $\beta$), as well as
nonlinear dispersion (characterized by the coefficient $\mu$). The latter is
known as ``self-steepening'' term, due to its effect on
the pulse envelope. 
These conservative higher-order terms (including the one of strength $\nu$) are naturally important not only in optics, but
also in the contexts of nonlinear metamaterials \cite{p31,p32,p33} and water waves 
\cite{johnson,sedletsky,slun}.
All the above effects, which are not incorporated in
the standard NLS equation ($\gamma=\delta=\beta=\mu=\nu=\sigma_R=0$),
become important in the physically relevant case of ultrashort pulses.

On the other hand, and in the same nonlinear optics context,
for relatively long propagation distances, dissipative effects, such
as linear loss ($\gamma<0$) [or gain ($\gamma>0$)], as well as
stimulated Raman scattering 
(of strength $\sigma_R>0$), are also included \cite{KodHas87,Agra1,Agra2}. At the same time,
it is also relevant to include
the nonlinear gain ($\delta>0$) [or loss ($\delta<0$)]
to counterbalance the effects from the linear loss/gain mechanisms \cite{p35}.
For relevant physical settings, the role of such terms is to contribute to the possible stabilization of solitons.
Notice that such solutions exist in the standard NLS equation, which
possesses solitons of two different types:
bright solitons (for $s=-1$) that decay to zero at infinity, and dark solitons (for $s=+1$) that
have the form of density dips on top of a continuous-wave (cw) background accompanied by a phase
jump at the density minimum. Below we focus on the latter case of normal dispersion ($s=+1$), for which
the linear/nonlinear dissipative effects usually lead to the decay
of the cw background; this way, the composite object, composed by the background and the soliton, is destroyed.

Dark solitons were first predicted to occur as early as in the early 70's \cite{suz71,Has73}, and since then have been
studied extensively, both in theory and in experiments, mainly in nonlinear optics \cite{Kiv98} and
Bose--Einstein condensation \cite{KFG1,KFG2,DJF}.
Concerning the higher-order NLS (\ref{eq1}), 
in Ref.~\cite{DJFTheo} it was shown, via a newly developed perturbation theory \cite{MJA}, that 
dark and grey solitons can exist and feature a stable evolution --at least for certain parameter regimes
and time intervals, within the validity of the perturbation theory. However, their long-time dynamics are under question.
The results of Ref.~\cite{DJFTheo} suggest that these dynamics, and different scenarios corresponding to
various parameter regimes, can be very rich.


The structure of the presentation and main findings of the present work are as follows.
In Sec.~\ref{SECII}, we explore the
finite-time collapse for the solutions of Eqs.~(\ref{eq1})--(\ref{eq3}). Our analysis is motivated from
the results and questions posed on the dynamics of the soliton background discussed in Ref.~\cite{DJFTheo}.
The discussion therein, shows that there can be a finite-time for which the background exhibits blow-up.
Also in Section \ref{SECIIa}, based on 
energy arguments carried out on a spatially averaged
$L^2$-energy functional, we identify that the collapse behavior of the considered $L^2$-functional
is chiefly characterized by the gain/loss parameters $\gamma$ and $\delta$.
Regimes of collapse and decay of solutions are defined by $\gamma,\delta>0$ and $\gamma,\delta<0$.
However, for
$\gamma<0,\,\delta>0$, 
we obtain
the existence of a critical value on the linear gain.
This critical value, say $\gamma^*$, is
separating finite-time collapse from the decay of solutions.
An important conclusion regarding the regimes $\gamma,\delta>0$ and $\gamma,\delta<0$
is related to the fate of localized solutions,
with an initial profile resembling that
of bright and dark solitons: referring to their
long-time stability, such wave forms are unstable, and they may survive
only for relevant short times.

The above picture on the asymptotic behavior is verified completely by the results of the numerical
simulations presented in Section \ref{SECIIb}.
We examine the evolution for three distinct types of initial conditions, including decaying and non-decaying data.
In the case of cw initial conditions, we observe that the analytical upper bound is not only fulfilled,
but also saturated by the analytical blow-up times. Furthermore, the  numerical collapse time is
independent of the strengths of the rest of the dissipative/conservative higher-order terms.
The observed saturation is completely justified by the fact that the timescales for collapse
of the background, and the analytical upper bound for the collapse of the $L^2$ functional,
coincide in the case of the cw initial condition. Also, the evolution of the $L^2$-functional
is proved to be that of the cw-background. Concerning the critical value $\gamma^*$, it is verified
numerically that its analytical estimation is very accurate; 
for $\gamma<\gamma^*$ the solution decays and
for $\gamma\geq\gamma^*$ the solution blows-up in finite-time.
In the case of the decaying
initial data (which are taken in the form of a $\mathrm{sech}$ function --
as reminiscent of a bright soliton of the focusing NLS), we have observed very interesting transient
dynamics prior to collapse when $\gamma\geq\gamma*$,
or decay when $\gamma<\gamma^*$, characterized by a self-similar evolution of the initial pulse for small values
of the parameters. As before, the analytical upper bound for the blow-up time is in agreement with
the numerical one. However, it is not saturated since the evolution of the $L^2$ functional is dominated
by the evolution of the amplitude of the localized solution. We have also tested numerically the behavior
of the blow-up time with respect to large variations of the strengths of linear and nonlinear parameters 
$\beta$ and $\sigma_R$, and found the following decreasing step function effect:
the numerical blow-up time jumps to a smaller numerical value but remains constant
within specific regimes of variation for the increasing parameters. The jump is associated with
a transition to a different type of transient dynamics prior to collapse or decay.

The numerical results were also found to be in excellent agreement with the analytical ones
in the case of the non-decaying initial data --taken here in the form of a $\mathrm{tanh}^2$ function,
which resembles the intensity profile
of a dark soliton of the defocusing NLS, and is
compliant with the periodic boundary conditions (\ref{eq2}).
The numerical blow-up time 
is in agreement with --and is also found to almost saturate to-- its analytical upper bound,
as in the case of the cw-initial condition, at least for small variations of the parameters,
for reasons that will be discussed later in the text.
In regimes of
larger parameter values,
we observe again the decreasing step function effect for the numerical blow-up time
associated with a drastic change in the transient dynamics.
In the case $\gamma<0,\,\delta>0$, it appears that the condition $\gamma\delta<0$ is not sufficient to define a stabilization regime with nontrivial dynamics or avoiding collapse. It will be shown in a subsequent work \cite{PartII}, that such a stabilization regime associated with emergent dynamics different from collapse or decay, may be defined only when $\gamma>0$, $\delta<0$.

\section{Conditions for collapse
}
\label{SECII}
\subsection{Collapse in finite-time for an $L^2$-norm functional}
\label{SECIIa}
The  question of collapse concerns sufficiently smooth (weak) solutions of equation (\ref{eq1}).
The existence of such solutions, is guaranteed by the following local existence result associated with the initial-boundary value problem (\ref{eq1})-(\ref{eq3}).
The methods which are
used in order to prove such local existence result in the Sobolev spaces of periodic functions $H^k_{per}$ \cite{Caz03,Sulem99}, are based on the lines of approach of \cite{Kato0}-\cite{Kato3}. The application of these methods to the problem (\ref{eq1})-(\ref{eq3}), although involving lengthy computations, is now considered as standard. Thus, we refrain
from showing  the details here, and we just state it in:

\setcounter{theorem}{0}
\begin{theorem}\label{thmloc}
	Let $u_0\in H^k_{per}(\Omega)$ for any integer $k\geq2$, and $\beta,\gamma,\delta,\mu,\nu, \sigma_R\in\mathbb{R}.$
	Then there exists $T>0$, such that the problem
	(\ref{eq1})-(\ref{eq3}), has a unique
	solution
	\begin{equation*}
	u\in C([0,T], H^k_{per}(\Omega)) \quad\mbox{and}\quad u_t\in
	C([0,T], H^{k-3}_{per}(\Omega)).
	\end{equation*}
	Moreover, the solution $u\in H^k_{per}(\Omega)$ depends continuously on the initial data $u_0\in H^k_{per}(\Omega)$, i.e., the solution operator
	\begin{eqnarray}
	\label{DSa}
	\mathcal{S}(t): H^k_{per}(\Omega)&\mapsto&  H^k_{per}(\Omega),\;\;\;\;t\in[0,T],\\
	u_0&\mapsto&\mathcal{S}(t)u_0=u,\nonumber
	\end{eqnarray}	
	is continuous.
\end{theorem}
Here, $H^k_{per}(\Omega)$ denote the Sobolev spaces of $2L$- periodic functions on the fundamental interval $\Omega=[-L,L]$. Let us recall for the shake of completeness their definition: 
\begin{eqnarray}
\label{defSob}
H^k_{per}(\Omega)&=&\{u:\Omega\rightarrow \mathbb{C},\;\; u\;\mbox{and}\; \frac{\partial^ju}{\partial x^j}\in L^2(\Omega),\;\; j=1,2,...,k;\nonumber\\
&&u(x),\;\;\mbox{and}\;\;\frac{\partial^ju}{\partial x^j}(x)\;\mbox{for $j=1,2,...,k-1$, are $2L$-periodic}
\}.
\end{eqnarray}
Since our analytical energy arguments for examining collapse require sufficiently smoothness of local-in-time solutions, we shall implement Theorem~\ref{thmloc} by assuming that $k=3$, at least. As it follows from the definition of the Sobolev spaces (\ref{defSob}), this assumption means that the initial condition $u_0(x)$, $x\in\Omega$, and its spatial (weak) derivatives, at least up to the 2nd-order, are $2L$-periodic. Then, it turns out from Theorem~\ref{thmloc},  that the unique, local-in-time solution $u(x,t)$ of (\ref{eq1}) satisfies the periodic boundary conditions (\ref{eq2}) for $t\geq 0$, and is sufficiently (weakly) smooth. Such periodicity and smoothness properties of the local-in-time solutions are sufficient for our purposes (see Theorem~\ref{T1a} and Appendix A, below). 

Next, we adopt the method of deriving a differential inequality for the functional
\begin{eqnarray}
\label{eq4a}
M(t)=\frac{{{e}^{-2\gamma t}}}{2L }\int_{-L}^{L}|u(x,t)|^2dx,
\end{eqnarray}
and then, showing that its solution diverges in finite-time under appropriate assumptions
on its initial value at time $t=0$; see \cite{Caz03, Ball77, Sulem99,OY03,TaylorII} and references therein.
Note that the choice of this functional is not arbitrary; in fact, it is a direct consequence of the conservation laws
of the NLS model (see 
Appendix A).
For a discrete counterpart  of this argument applied in discrete Ginzburg-Landau-type equations, we refer to \cite{NET07}. For applications of these types of arguments in the study of escape dynamics for Klein-Gordon chains, we refer to \cite{V13}.
\begin{theorem}
	\label{T1a}
	For $u_0\in  H^k_{per}(\Omega)$, $k\geq3$, let $\mathcal{S}(t)u_0=u\in C([0,T_{\mathrm{max}}), H^k_{per}(\Omega))$ be the local- in- time solution of the problem (\ref{eq1})-(\ref{eq3}), with  $[0,T_{\mathrm{max}})$ be its maximal interval of existence. Assume that the parameter $\delta>0$ and that the initial condition  $u_0(x)$ is such that $M(0)>0$. Then, $T_{\mathrm{max}}$ is finite, in the following cases:
	\begin{eqnarray}
	\label{a1a}
	&&(i)\;\;\;\;\;\;\;\;\;\;\;{{T}_{\mathrm{max}}}\le \frac{1}{2\gamma }\log \left[ 1+\frac{\gamma }{\delta M\left( 0 \right)} \right],\\
	\label{a2a}
	&&\mbox{for}\;\;\gamma\neq 0,\;\;\mbox{and}\;\;\
\gamma>-\delta M\left( 0 \right).\\
	\label{b1a}
	&&(ii)\;\;\;\;\;\;\;\;\;\;\;{{T}_{\mathrm{max}}}\le \frac{1}{2\delta M\left( 0 \right)},\;\;\mbox{for}\;\;\gamma=0.
	\end{eqnarray}
\end{theorem}
{\textbf{Proof:} For any $T<T_{\mathrm{max}}$, since $k\geq 3$,  due to the continuous embedding $C([0,T], H^k_{per}(\Omega))\subset C([0,T], L^2(\Omega))$ \cite{Caz03}, the solution $\mathcal{S}(t)u_0=u\in C([0,T], L^2(\Omega))$. Furthermore, it follows from Theorem \ref{thmloc}, that $u_t\in
C([0,T], L^2(\Omega))$. Then, by differentiating $M(t)$ with respect to time, we find that
	\begin{eqnarray}
	\label{OL1}
	\frac{dM(t)}{dt}=-\gamma \frac{e^{-2\gamma t}}{L}\int_{-L}^{L}|u|^2dx
	+\frac{e^{-2\gamma t}}{L}\mathrm{Re}\int_{-L}^{L}u_t\overline{u}dx.
	\end{eqnarray}
In the second 
term on the right-hand side of (\ref{OL1}), we substitute $u_t$ by the right-hand side of Eq.~(\ref{eq1}).
Then, after some computations (see details in Appendix A), Eq.~(\ref{OL1}) results in:
	\begin{eqnarray}
	\label{OL1N}
	\frac{dM(t)}{dt}=\delta \frac{e^{-2\gamma t}}{L}\int_{-L}^{L}|u|^4dx.
	\end{eqnarray}
	Next, by the Cauchy-Schwarz inequality, we have
	\begin{eqnarray}
	\label{CS}
	\int_{-L}^{L}|u|^2dx\leq \sqrt{2L}\left(\int_{-L}^{L}|u|^4dx\right)^{\frac{1}{2}}.
	\end{eqnarray}
	Therefore, for the functional $M(t)$ defined in (\ref{eq4a}), we get the inequality
	\begin{eqnarray}
	\label{OLe1}
	M(t)\leq \frac{e^{-2\gamma t}}{\sqrt{2L}}\left(\int_{-L}^{L}|u|^4dx\right)^{\frac{1}{2}},
	\end{eqnarray}
	which in turns, implies the estimate
	\begin{eqnarray}
	\label{OLe1es}
	M(t)^2\leq \frac{e^{-4\gamma t}}{2L}\int_{-L}^{L}|u|^4dx,
	\end{eqnarray}
	for all $t\in[0, T_{\mathrm{max}})$. On the other hand,  from (\ref{OL1N}) we have that
	\begin{eqnarray*}
		\int_{-L}^{L}|u|^4dx=e^{2\gamma t}\frac{L}{\delta}\,\frac{dM(t)}{dt},
	\end{eqnarray*}
	and hence, we may rewrite (\ref{OLe1es}) as
	\begin{eqnarray}
	\label{OLe2}
	{{\left[ M\left( t \right) \right]}^{2}}\le \frac{{{e}^{-2\gamma t}}}{2\delta }\frac{dM\left( t \right)}{dt},\;\;\mbox{or}\;\;
	\frac{\frac{dM\left( t \right)}{dt}}{{{\left[ M\left( t \right) \right]}^{2}}}\ge 2\delta {{e}^{2\gamma t}}.
	\end{eqnarray}
	Since $\frac{d}{dt}\left[ \frac{1}{M\left( t \right)} \right]=-\frac{\frac{dM\left( t \right)}{dt}}{{{\left[ M\left( t \right) \right]}^{2}}}$,  we get from (\ref{OLe2}) the differential inequality
	\begin{eqnarray}
	\label{OLe3}
	\frac{d}{dt}\left[ \frac{1}{M\left( t \right)} \right]\le -2\delta {{e}^{2\gamma t}}.
	\end{eqnarray}
	Integration of (\ref{OLe3}) with respect to time, implies that
	\begin{eqnarray*}
		\frac{1}{M\left( t \right)}\le \frac{1}{M\left( 0 \right)}-2\delta \int_{0}^{t}{{{e}^{2\gamma s}}ds}.
	\end{eqnarray*}
	and since $M(t)>0$, we see that $M(0)>0$ satisfies the inequality
	\begin{eqnarray}
	\label{OLe4}
	2\delta \int_{0}^{t}{{{e}^{2\gamma s}}ds}\le \frac{1}{M\left( 0 \right)}.
	\end{eqnarray}
	From (\ref{OLe4}), we shall distinguish between the following cases for the damping parameter $\gamma$:
	%
\begin{itemize}
\item
We assume that the damping parameter $\gamma\neq 0$. In this case, (\ref{OLe4}) implies that
	\begin{eqnarray*}
		\frac{2\delta }{2\gamma }\left( {{e}^{2\gamma t}}-1 \right)\le \frac{1}{M\left( 0 \right)},\;\;\mbox{or}\;\;
		{{e}^{2\gamma t}}\le 1+\frac{\gamma }{\delta M\left( 0 \right)}.
	\end{eqnarray*}
	Thus, assuming that $\frac{\gamma }{\delta M\left( 0 \right)}>-1$, we derive that the maximal time of existence is finite, since
	\begin{eqnarray*}
		t \le \frac{1}{2\gamma }\log \left[ 1+\frac{\gamma }{\delta M\left( 0 \right)} \right].
	\end{eqnarray*}
	The inequality above, proves the estimate of the collapse time (\ref{a1a}) under assumption (\ref{a2a}), that is, case (i) of the Theorem.
\item 	
	Assume that the damping parameter is $\gamma=0$. 
	Then, Eq.~(\ref{OLe4}) implies that
	$2\delta t\le \frac{1}{M\left( 0 \right)}$, or
	\begin{eqnarray*}
		t\le \frac{1}{2\delta M\left( 0 \right)}.
	\end{eqnarray*}
	This inequality proves the estimate of the collapse time (\ref{b1a}) in the absence of damping, that is,
	case (ii) of the Theorem.\;\;\;\;$\Box$
\end{itemize}	
	From condition
	(\ref{a2a}) on the definition of the analytical upper bound of the blow-up time
	\begin{eqnarray}
	\label{eqUB}
	T_{\mathrm{max}}[\gamma,\delta,M(0)]=\frac{1}{2\gamma }\log \left[ 1+\frac{\gamma }{\delta M\left( 0 \right)} \right],
	\end{eqnarray}	
	given in (\ref{a1a}), we define a critical value of the linear gain/loss parameter as
	\begin{eqnarray}
	\label{CRIT}
	\gamma^{*}=-\delta M(0).
	\end{eqnarray}
	We observe that
	\begin{eqnarray}
	\label{eqUB2}	
	\lim_{\gamma\rightarrow\gamma^*}T_{\mathrm{max}}[\gamma,\delta,M(0)]=+\infty,
	\end{eqnarray}
	while $T_{\mathrm{max}}[\gamma,\delta,M(0)]$ is finite if
	\begin{eqnarray}
	\label{eqUB2a}
	\gamma>\gamma^*,
	\end{eqnarray}
	according to the condition (\ref{a2a}). Then, (\ref{eqUB2}) suggests that when $\delta>0$, the critical 
	value $\gamma^*$ may act as a critical
point separating the two dynamical behaviors:  blow-up in finite-time for $\gamma>\gamma^*$ and global existence for $\gamma<\gamma^*$.  We shall investigate this  scenario numerically.
	
	We also remark that the analytical upper bound for the blow-up time (\ref{b1a}) in the case $\gamma=0$,
	\begin{eqnarray}
	\label{eqUB0}
	T_{\mathrm{max}}[\delta,M(0)]=\frac{1}{2\delta M\left( 0 \right)},
	\end{eqnarray}
	is actually the limit of the analytical upper bound (\ref{eqUB}) for $\gamma>0$ as $\gamma\rightarrow 0$, e.g.,
	\begin{eqnarray}
	\label{eqUB01}	
	\lim_{\gamma\rightarrow 0}T_{\mathrm{max}}[\gamma,\delta,M(0)]=T_{\mathrm{max}}[\delta,M(0)].
	\end{eqnarray}

\section{Conditions for collapse in finite-time: numerical results}
\label{SECIIb}
In this section, we verify numerically the collapse of different types of
initial data.
We shall study three particular cases for the initial data: (a) 
cw initial conditions, (b) $\mathrm{sech}$-profiled initial conditions, resembling a ``bright soliton'',
and (c) $\mathrm{tanh}^2$-profiled initial conditions, resembling a ``dark soliton''. 

All numerical experiments have been performed using a high accuracy pseudo-spectral method, incorporating an embedded error estimator that makes it possible to efficiently determine appropriate step sizes. We denote as time of collapse the time where the numerical scheme detects a singularity in the dynamics. The numerical singularity is detected when the time step is becoming of order $<10^{-12}$.
%
Concerning the implementation of the periodic boundary conditions for 
cases (b) and (c) of the initial data, let us also recall that in both cases, 
these conditions are strictly satisfied only asymptotically, as $L\rightarrow\infty$; 
for a finite length $L$, these initial profiles, as well as their spatial derivatives, 
have jumps across the end points of the interval $[-L, L]$. However, these jumps have 
negligible effects in the observed dynamics, as they are of order $\exp(-L)$. More precisely, 
since the smallest value for $L$ used herein is $L=20$, the derivatives are 
O$(10^{-9})$, 
or less.

\subsection{Continuous wave 
initial conditions.}
\label{CA}
The first example for our numerical study, concerns cw initial data, of the form:
\begin{eqnarray}
\label{pw1}
u_0(x)=\epsilon\, e^{-\mathrm{i}\frac{K\pi x}{L}},\;K>0,
\end{eqnarray}
of amplitude $\epsilon>0$ and wave-number $K>0$. 
This type of plane wave dynamics can be understood
by means of the substitution in Eq.~(\ref{eq1}) for ansatz of the
form $u(x,t)=W(t) e^{\mathrm{i} K x}$, in line with the above presented
numerical simulations. Then the resulting dynamics is:
\begin{eqnarray}
W_t-\mathrm{i} \frac{s K^2}{2} W- \mathrm{i} |W|^2 W = \gamma W + \delta |W|^2 W
- \mathrm{i} \beta K^3 W + \mathrm{i} \mu K W
\label{extra1}
\end{eqnarray}
Subsequent introduction of a phase factor in the form of
$W(t)=e^{\mathrm{i} \omega t} w(t)$ with $\omega=s K^2/2 - \beta K^3 + \mu K$
absorbs through this gauge transformation a number of the terms
in Eq.~(\ref{extra1}) finally leading to:
\begin{eqnarray*}
\mathrm{i}\dot{w}-|w|^2w=\mathrm{i}\gamma w+\mathrm{i}\delta|w|^2w,
\end{eqnarray*}
where $\dot{w}=dw/dt$.
Then, employing the polar decomposition $w=h(t)e^{\mathrm{i}\theta(t)}$ we get
\begin{eqnarray}
\label{ODE1}
\dot{h}=\gamma h+\delta h^3.
\end{eqnarray}
Assuming that the initial height of the wave background is $h(0)=h_0$, the ODE (\ref{ODE1}) has the general solution
\begin{eqnarray}
\label{ODE2}
h^2(t)=\frac{\gamma h_0^2e^{2\gamma t}}{\gamma+\delta h_0^2-\delta h_0^2e^{2\gamma t}}.
\end{eqnarray}
The solution of (\ref{ODE2}) blows-up in the finite-time
\begin{eqnarray}
\label{ODE3}
T^*_{\mathrm{max}}=\frac{1}{2\gamma }\log \left[ 1+\frac{\gamma}{\delta h_0^2} \right],
\end{eqnarray}
exactly under the same assumptions on the strengths $\gamma$ and $\delta$ stated in Theorem \ref{T1a}.

Recalling the above analysis, for the initial data (\ref{pw1}) we have
\begin{eqnarray*}
	M(0)=\frac{1}{2L}\int_{-L}^{L}\epsilon^2|e^{-\mathrm{i}\frac{K\pi x}{L}}|^2dx=\epsilon^2.
\end{eqnarray*}
Theorem \ref{T1a} and (\ref{eqUB})  assert that the analytical upper bound of the collapse time $T_{\mathrm{max}}[\gamma,\delta, M(0)]$ is
\begin{eqnarray}
\label{ap1I}
T_{\mathrm{max}}[\gamma,\delta,\epsilon]=\frac{1}{2\gamma }\log \left[ 1+\frac{\gamma}{\delta\epsilon^2} \right].
\end{eqnarray}
According to (\ref{CRIT}), the critical value $\gamma^*$ in this case 
[see also (\ref{ap1I})] is:
\begin{eqnarray}
\label{ap2I}
\gamma^*=-\delta\epsilon^2.
\end{eqnarray}
Also, (\ref{eqUB0}) asserts that the analytical upper bound for the collapse time in the case $\gamma=0$ is
\begin{eqnarray}
\label{ap4I}
T_{\mathrm{max}}[\delta,\epsilon]=\frac{1}{2\delta \epsilon^2}.
\end{eqnarray}

\begin{figure}
	\begin{center}
		\begin{tabular}{c}
			\includegraphics[scale=0.41]{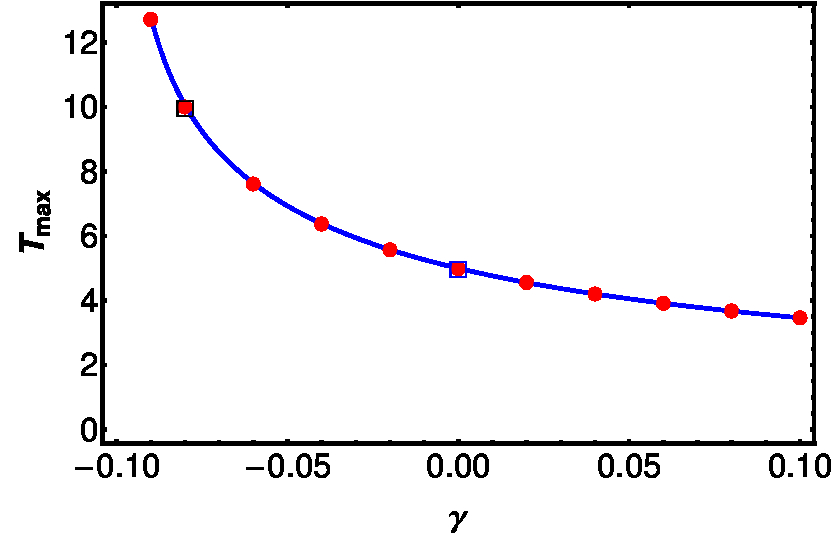}
			\includegraphics[scale=0.4]{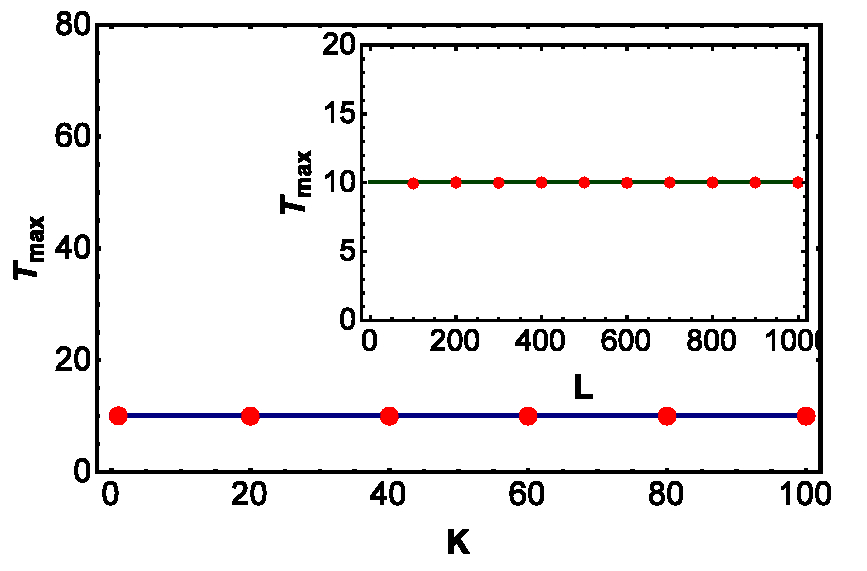}
		\end{tabular}
		\caption{(Color Online) 
		Left panel: Numerical blow-up time for the plane wave initial data (\ref{pw1}) against its analytical upper bound (\ref{ap1I}), as a function of $\gamma\in [-0.09, 0.1]$, for 
		$\delta=0.1$, $\mu=\nu=0.01$, $\sigma_R=0.02$, $\beta=1$. Continuous (blue) curve 
		depicts the analytical upper bound and (red) dots 
		the numerical blow-up time. Here $L=100$, while 
		$K=10$, $\epsilon=1$ for the cw. The analytical upper bound for 
		$\gamma=0$ is marked by the (blue) square. 
		Right panel: Independence of the blow-up time on the wave number $K$ for cw
		initial data and on the length $L$ of the spatial interval (inset). Here, $\gamma=-0.08$, the rest of
		parameters are 
		as above, and for fixed $L=100$, $K\in [10,100]$ (main panel), while for fixed $K=10$, $L\in [100,1000]$ (inset). For this set of parameters, the analytical upper bound for the blow-up time is $T_{\mathrm{max}}=10.05$ (lines-marked by the black square in the upper figure), while the numerical blow-up time is  $T_{\mathrm{num}}\approx 10.01$ (dots).}
		\label{fig1}
	\end{center}
\end{figure}

The left panel of Fig.~\ref{fig1}, illustrates the comparison of the analytical upper bound on the collapse time (\ref{ap1I}), against its numerical value, as a function of the parameter $\gamma$, fixing the rest of parameters at specific examples of values. We observe that the analytical upper bound seems to be a very accurate estimate of the numerical collapse time, and is almost saturated.  This, one would argue, is not surprising provided that the solution preserves
its plane wave structure, since the latter indeed saturates the
Cauchy-Schwarz inequality (\ref{CS})
used to derive our differential inequality
from which the collapse time was obtained.
Indicative of the accuracy of the analytical estimate is the comparison between the analytical upper bound for the case $\gamma=0$, and the numerical blow-up time for the considered set of parameters: the analytical value (\ref{ap4I}) is $T_{\mathrm{max}}[0.1,1]$=5 (marked by the blue square in the figure), and the numerical blow-up time (the red dot inside the square) is $T_{\mathrm{num}}=4.99$.
The analytical upper bound (\ref{ap1I}) does not depend on
the wavenumber $K$ and the half-length $L$ of the symmetric spatial interval $[-L,L]$. This independence is also preserved by the numerical blow-up time as illustrated in the right panel of Fig.~\ref{fig1}. The observed independence
is justified by the same reasons as above
(i.e., that the inequality of~(\ref{CS}) is indeed an equality for
this type of solutions, provided they maintain their spatial form).

Figure~\ref{fig2}, illustrates the comparison of the analytical upper bound on the collapse time (\ref{ap1I}) against the numerical collapse time, as a function of the parameter $\delta>0$, the amplitude of the plane wave $\epsilon$, and the  parameters $\sigma_R$, $\beta$, $\mu$, and $\nu$. In the upper panel, we observe again the high accuracy of the analytical upper bound for the collapse time (\ref{ap1I}), as a function of the nonlinear gain/loss parameter $\delta$, and as a function of the amplitude $\epsilon$. In the bottom panel, we illustrate the independence of the  numerical blow-up time from the parameters $\sigma_R$, $\beta$, $\mu$ and $\nu$, as predicted by the analytical estimate.
We remark that the last two terms of the nonlinearity of (\ref{eq1}) can been written as
\begin{eqnarray*}
	F[u]=[(\mu+\nu)-\mathrm{i}\sigma_R]|u|^2u_x
	+[(\mu+\nu)-\mathrm{i}\sigma_R]u^2\overline{u}_x.
\end{eqnarray*}
Thus, we have integrated the system varying the sum $\mu+\nu$.
 \begin{figure}
 	\begin{center}
 		\begin{tabular}{c}
 			\includegraphics[scale=0.4]{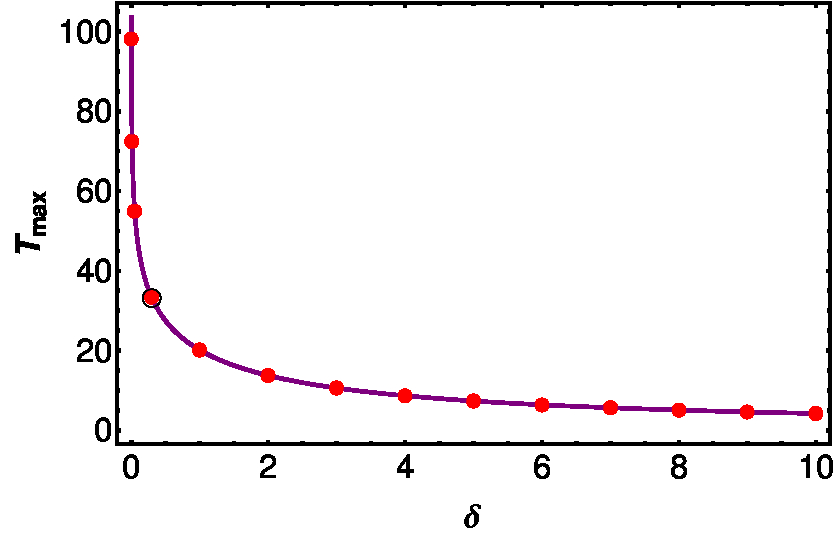}
 			\includegraphics[scale=0.4]{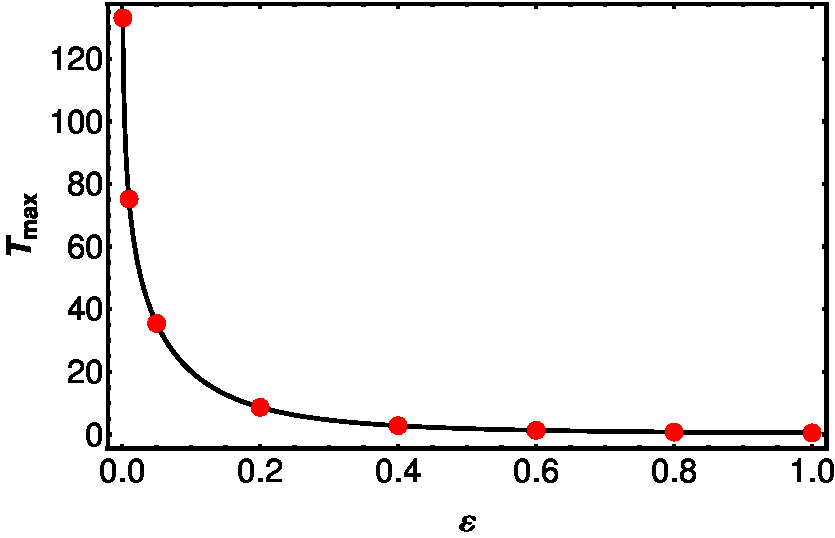}\\
 			\includegraphics[scale=0.4]{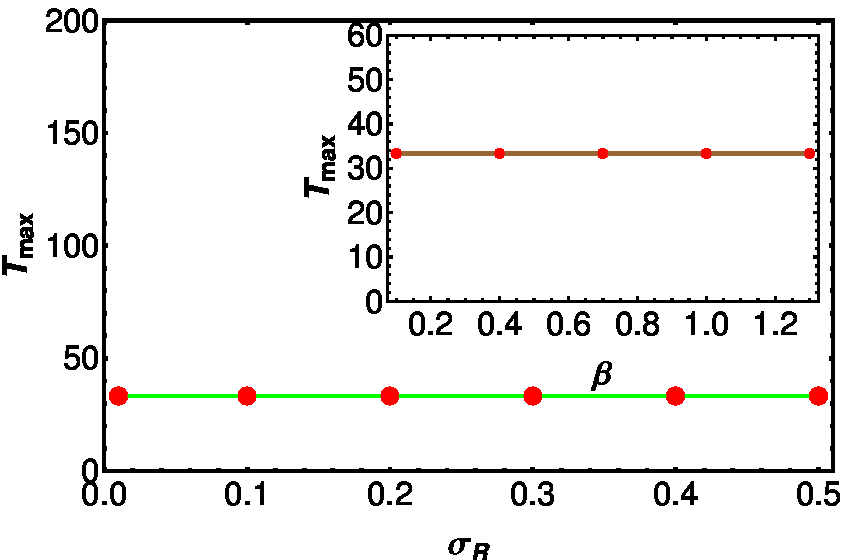}
 			\includegraphics[scale=0.4]{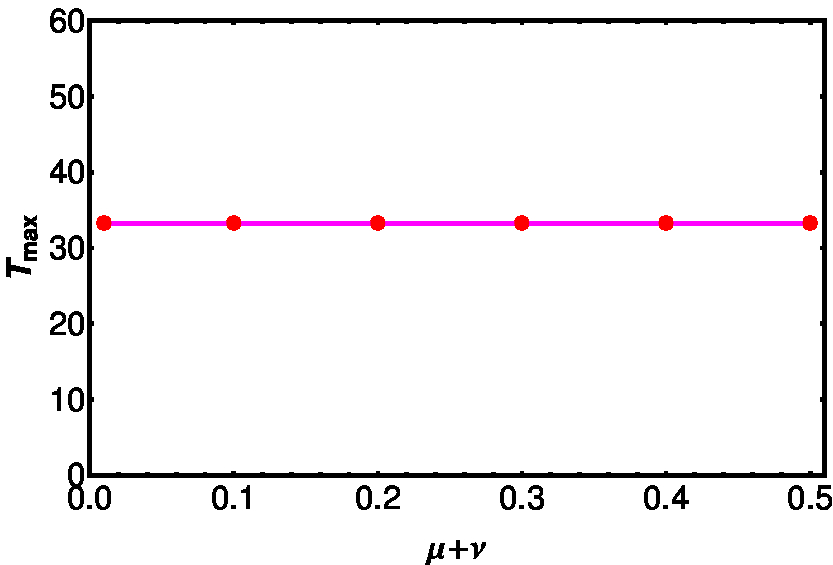}
 		\end{tabular}
 		\caption{(Color Online) 
 			Top left panel: Numerical blow-up time for the
 			cw initial data (\ref{pw1}) against its analytical upper bound (\ref{ap1I}), as a function
 			of $\delta\in [0.001, 10]$, for 
 			$\gamma=0.04$, $\mu=\nu=0.01$, $\sigma_R=0.02$, $\beta=1$.
 			The continuous (purple) curve
 			corresponds to the analytical upper bound (\ref{ap1I}) and (red) dots to the numerical blow-up time.
 			Here, $L=100$, while 
 			$K=10$, $\epsilon=0.1$ for the cw.
 			Top right panel: Numerical blow-up time (dots) against the analytical upper bound (continuous (black) curve) as a function of $\epsilon\in [0.001,1] $, for 
 			$\gamma=0.04$, $\delta=1$, $\mu=\nu=0.01$, $\sigma_R=0.02$, $\beta=1$, $L=100$, $K=10$. 
 			Bottom left panel: Independence of the blow-up time on the parameters $\sigma_R$ and $\beta$ (inset). Here, $\gamma=0.04$, $\delta=0.3$, $\epsilon=0.1$, $K=10$, $L=100$; the
 			analytical upper bound for the blow-up time (black circles)
 			is $T_{\mathrm{max}}=33.28$.
 			(lines), is marked by the black circle in the top
 			panels.
 			In the main panel, 
 			$\beta=1$, and $\sigma_R\in [0.01,0.5]$, while 
 			$\sigma_R=0.3$, $\beta\in [0.1,1.3]$ in the inset. The numerical blow-up time is
 			$T_{\mathrm{num}}\approx 33.21$ (dots). Bottom right panel: Independence of the blow-up time on 
 			parameters $\mu+\nu\in [0.01,0.5]$, for $\beta=1$, $\sigma_R=0.3$; 
 			the rest of parameters are as 
 			in the upper right panel. }
 		\label{fig2}
 	\end{center}
 \end{figure}

In Fig.~\ref{fig3}, we present the results of a numerical study testing the behavior of the critical value (\ref{ap2I}) for the linear gain/loss parameter $\gamma$, as a separatrix between global existence and collapse. The left panel shows the evolution of the density for plane wave initial data (\ref{pw1}), for the parameters example of the upper panel of Fig. \ref{fig1}, and the choice of the parameter $\gamma$ exactly at the critical value, $\gamma=\gamma^*=-0.1$.  We observe the blow-up of the density, found numerically at time $T_{\mathrm{num}}=27.56$. The collapse is similar to that of an unstable cubic ordinary differential equation (ODE), with a rapid, spatially uniform, monotonic increase of the density as the evolution approaches the collapse time.
Notice the preservation (both in this case and in the case discussed
below) of the plane wave character of the solution.
On the other hand, slightly perturbing the parameter $\gamma<\gamma^*$, we observe the complete change in the asymptotic behavior, exhibiting the  decay of solutions. This is shown in the right panel, illustrating the evolution of the density for plane wave initial data for the parameters example of the upper panel, but for $\gamma-0.101<\gamma^*=-0.1$, slightly below from the critical value $\gamma^*$. The decay is again spatially uniform, resembling this time, that of a stable cubic ODE. We observe that the time-scales for blow-up (\ref{eqUB}) ---and (\ref{ap1I}) in the case of the plane wave initial data---
coincide with the time-scales for blow-up of the background $h^2$ (\ref{ODE3}).
Furthermore, it follows from the solution (\ref{ODE2}) that $\lim_{t\rightarrow\infty}h(t)^2=0$, if $\gamma<-\delta h_0^2$, for $\delta>0$, i.e., the condition for decay of the background $h(t)$ coincides with the condition $\gamma<\gamma^*$  for  the decay of the solution in the regime $\gamma<0,\,\delta>0$.
This coincidence explains the saturation of the analytical upper bound (\ref{ap1I}) by the numerical blow-up times from a complementary
perspective (to that of the saturation of the inequality of
Eq.~(\ref{CS})) and the overall agreement of the parametric dependencies
between the numerics and the theoretical analysis above.
\begin{figure}
	\begin{center}
		\begin{tabular}{c}
			\includegraphics[scale=0.46]{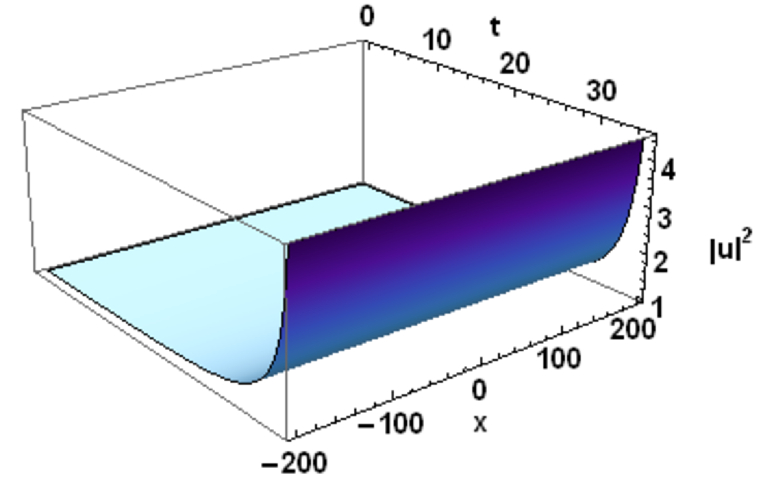}
			\includegraphics[scale=0.46]{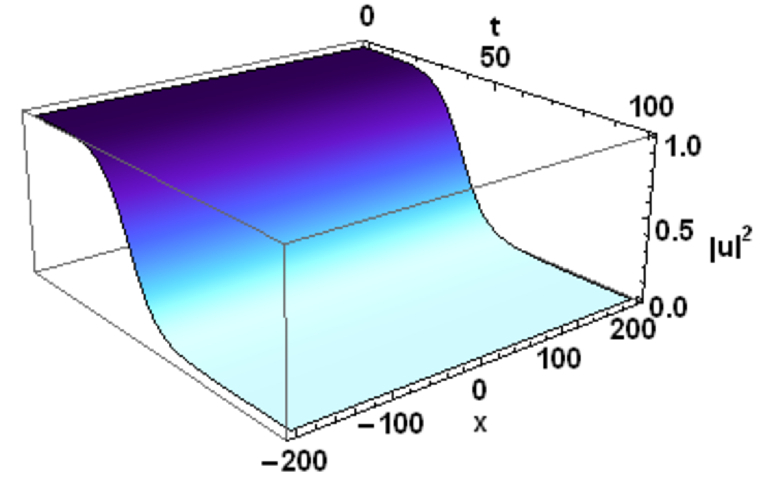}
		\end{tabular}
		\caption{(Color Online) Left Panel: Collapse of the density $|u|^2$ for plane-wave initial data (\ref{pw1}), as found for  $\delta=0.1$, $\epsilon=1$, and the critical value for $\gamma=\gamma^*-0.1$. The rest of parameters are $K=10$, $L=200$, $\mu=0.01$, $\nu=0.01$, $\sigma=0.02$, $\beta=1$. Right Panel: Decay of the density $|u|^2$ for plane-wave initial data, as found for  $\delta=0.1$, $\epsilon=1$, and  $\gamma=-0.101<\gamma^*-0.1$. The rest of parameters are the same as in the upper panel. }
		\label{fig3}
\end{center}
\end{figure}

\subsection{Decaying (sech-profiled) initial conditions.}
The second example for our numerical study concerns sech-profiled initial data
\begin{eqnarray}
\label{pw1a1}
u_0(x)=\epsilon\, \mathrm{sech} x,
\end{eqnarray}
of amplitude $\epsilon>0$. Such initial data correspond to
the profile of a ``bright soliton'' as an initial state. In this case,
\begin{eqnarray*}
	M(0)=\frac{1}{2L}\int_{-L}^{L}\epsilon^2\mathrm{sech}^2 xdx=\frac{\epsilon^2\mathrm{tanh}L}{L},
\end{eqnarray*}
and Theorem \ref{T1a} and (\ref{eqUB})  assert that the analytical upper bound of the collapse time $T_{\mathrm{max}}[\gamma,\delta, M(0)]$ is
\begin{eqnarray}
\label{ap1Ia1}
T_{\mathrm{max}}[\gamma,\delta,\epsilon,L]=\frac{1}{2\gamma }\log \left[ 1+\frac{\gamma L}{\delta\epsilon^2\mathrm{tanh}L} \right].
\end{eqnarray}
Also, (\ref{eqUB0}) asserts that the analytical upper bound for the collapse time in the case $\gamma=0$ is
\begin{eqnarray}
\label{ap4Ia1}
T_{\mathrm{max}}[\delta,\epsilon,L]=\frac{L}{2\delta \epsilon^2\mathrm{tanh}L}.
\end{eqnarray}
According to (\ref{CRIT}), the critical value $\gamma^*$ in this case
[cf. (\ref{ap1Ia1})] is:
\begin{eqnarray}
\label{ap2Ia1}
\gamma^*=-\frac{\delta\epsilon^2\mathrm{tanh}L}{L}.
\end{eqnarray}

Moreover, it follows from (\ref{ap2Ia1}), that for a sufficiently large
half-length parameter $L$,
\begin{eqnarray}
\label{ap4Icrit}
\gamma^*\approx 0.
\end{eqnarray}
\begin{figure}
	\begin{center}
		\begin{tabular}{c}
			\includegraphics[scale=0.42]{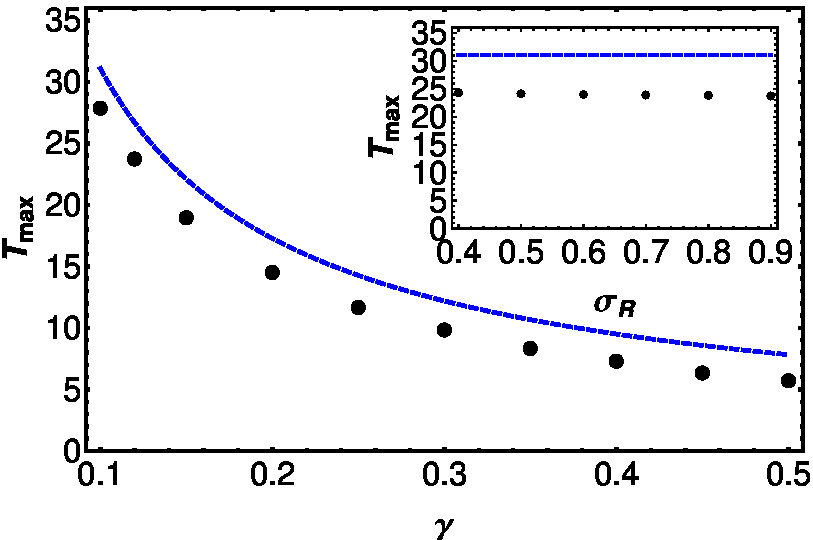}
			\includegraphics[scale=0.42]{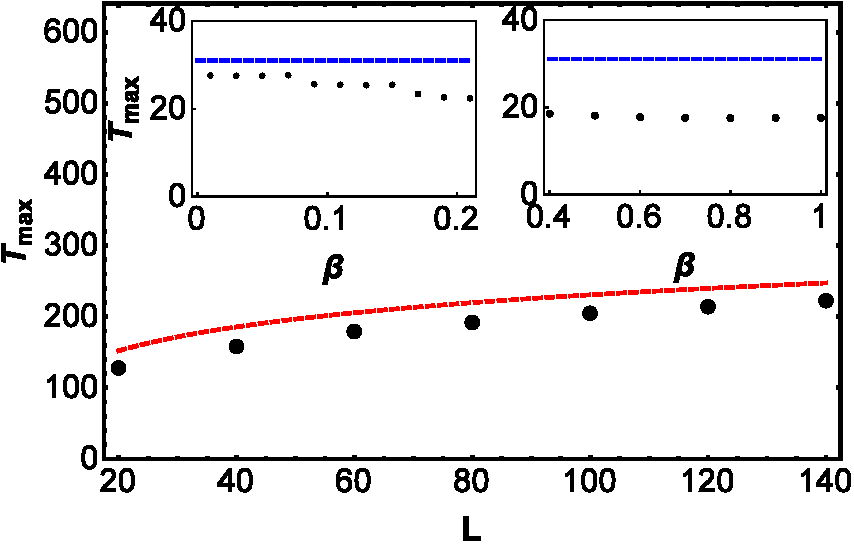}
		\end{tabular}
		\caption{(Color Online) Left panel: Numerical blow-up time for the $\mathrm{sech}$-profiled initial data (\ref{pw1a1}) against its analytical upper bound (\ref{ap1Ia1}), as a function of $\gamma\in [0.1, 0.5]$. Parameters $\delta=0.01$, $\mu=\nu=0.01$, $\sigma_R=0.02$, $\beta=0.02$. The dashed (blue) curve is traced for the analytical upper bound (\ref{ap1Ia1}) and (black) dots for the numerical blow-up time; here $L=50$. The inset illustrates the numerical blow-up time (dots) against the analytical upper bound (dashed (blue) curve) as a function of
			$\sigma_R\in [0.4,1] $. Parameters $\gamma=0.1$, $\delta=0.01$, $\mu=\nu=0.01$, $\sigma_R=0.02$, $\beta=0.02$, $L=50$. Right panel: Numerical blow-up time for the $\mathrm{sech}$-profiled initial data (\ref{pw1a1}) against its analytical upper bound (\ref{ap1Ia1}), as a function of $L\in [20, 140]$. Parameters $\gamma=0.01$, $\delta=0.01$, $\mu=\nu=0.01$, $\sigma_R=0.02$, $\beta=0.02$. The dashed (red) curve is traced for the analytical upper bound (\ref{ap1Ia1}) and (black) dots for the numerical blow-up time. The insets illustrate the numerical blow-up time (dots) against the analytical upper bound (dashed (blue) curve) as a function of
			$\beta\in [0.01,0.2]$ (left inset) and $\beta\in [0.4, 1]$ (right inset) .  Parameters $\gamma=0.1$, $\delta=0.01$, $\mu=\nu=0.01$, $\sigma_R=0.02$, $\beta=0.02$, $L=50$. }
		\label{fig4a}
	\end{center}
\end{figure}
\begin{figure}[h]
	\begin{center}
		\begin{tabular}{c}
			\includegraphics[scale=0.46]{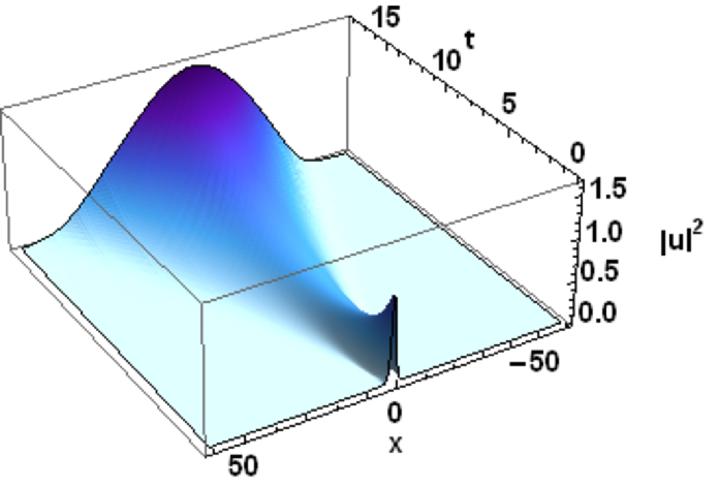}
			\includegraphics[scale=0.46]{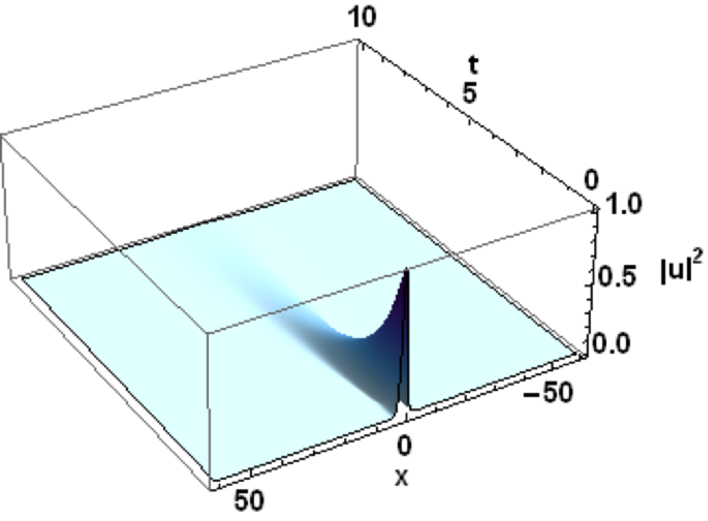}\\
			\hspace{-1cm}\includegraphics[scale=0.56]{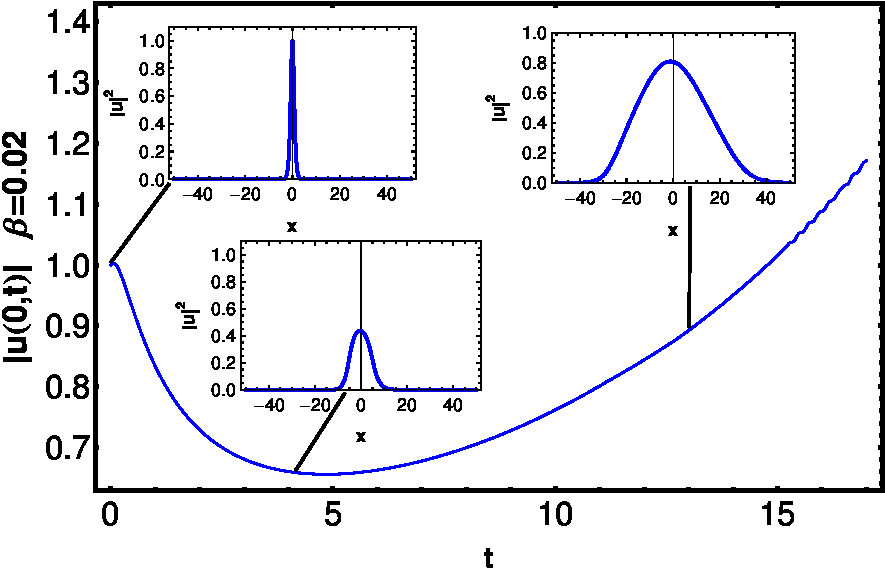}\hspace{-0.2cm}
		\end{tabular}
		\caption{(Color Online) Dynamics for $\mathrm{sech}$-type initial data,
			for $\beta=0.02$, 
			$\delta=0.01$, $\epsilon=1$, $\mu=0.01$, $\nu=0.01$, $\sigma=0.02$, and $L=50$.
			Upper left panel: Evolution of the density $|u|^2$ towards collapse, when $\gamma=0.1>\gamma^*\approx0$. Upper right panel: Decay of the density $|u|^2$ when $\gamma=-0.1<\gamma^*\approx 0$. Bottom panel: Evolution of $|u(0,t)|$ for the center of the pulse towards collapse when $\gamma=0.1>\gamma^*\approx0$, corresponding to the density evolution shown in the upper left panel. The insets demonstrate snapshots of the density $|u|^2$ for specific values of $|u(0,t)|$-marked by lines connecting the density profiles.}
		\label{fig5}
	\end{center}
\end{figure}
In the left panel of Fig.~\ref{fig4a}, we present the numerical study on the comparison of the numerical blow-up time for the $\mathrm{sech}$-profiled initial data (\ref{pw1a1}), against its analytical upper bound (\ref{ap1Ia1}), as a function of $\gamma\in [0.1, 0.5]$. The rest of parameters are $\delta=0.01$, $\mu=\nu=0.01$, $\sigma_R=0.02$, $\beta=0.02$ and $L=50$.
We observe that the analytical upper bound is again satisfied. However, it is not saturated as in the case of the
cw initial data (as discussed above).
Here, the time of collapse differs systematically by about $15$-$30$\%,
suggesting the potential for more interesting dynamical evolutions
that we now explore.
In fact, the case of the sech-profiled initial data, is much different since the initial data are spatially localized.
By continuity, the solution will be also spatially localized for a finite interval $[0,\tau]$ prior to collapse.
In this case, for the height of the wave-background we have $h(t)^2=0=\min_{x\in[-L,L]}|u(x,t)|^2$ ---the minimum of the density of the solution. Also,
in this case
we have the inequality
\begin{eqnarray}
\label{colcom2}
e^{-2\gamma t}h(t)^2&<&M\left( t \right)=\frac{{{e}^{-2\gamma t}}}{2L }\int_{-L}^{L}|u(x,t)|^2dx\nonumber\\
&<&\frac{{{e}^{-2\gamma t}}}{2L }\max_{x\in[-L,L]}|u(x,t)|^2\,2L\nonumber\\
&&=e^{-2\gamma t}\max_{x\in[-L,L]}|u(x,t)|^2,
\end{eqnarray}
satisfied by the amplitude $\max_{x\in[-L,L]}|u(x,t)|^2$ of the localized solution, for all $\;\;t\in[0,\tau]$. Thus, in the case of spatially localized data, we expect that the collapse will be predominately manifested as the increase of the amplitude, which grows faster than the averaged norm $M(t)$, due to (\ref{colcom2}). Consequently, the amplitude may be
expected to blow-up at an earlier time than $M(t)$.

It is also interesting to remark the slow logarithmic increase of the numerical blow-up time with respect to the parameter $L\in [20,140]$, as predicted by the analytical upper bound (\ref{ap1Ia1}), and shown in the right panel of Fig.~\ref{fig4a} i.e., although the quantitative collapse time may be different,
qualitatively our expression very accurately captures the relevant functional
dependences. 
The parameter values for this example are $\gamma=0.01$, $\delta=0.01$, $\mu=\nu=0.01$, $\sigma_R=0.02$, $\beta=0.02$.

The insets in both panels of Fig.~\ref{fig4a} examine numerically the independence of the collapse time on others among the rest of the nonlinear parameters.
In both examples we have considered the analytical upper bound $T_{\mathrm{max}}=31$, for
$\gamma=0.1$, $\delta=0.01$, $\mu=\nu=0.01$, $\sigma_R=0.02$, $\beta=0.02$, and $L=50$. The inset
in the left panel shows the small variations of the numerical blow-up time with respect to $\sigma_R\in [0.4,1]$,
around a mean constant value $T_{\mathrm{num,\sigma_R}}\approx 24$. The left inset in the right panel shows a "decreasing step function" effect
for the numerical blow-up time with respect to $\beta\in [0.01,0.2]$,  while the right inset shows that the numerical blow-up time has reached a mean constant value  $T_{\mathrm{num},\beta}\approx 17.8$ when $\beta\in[0.4,1]$.} The difference between $T_{\mathrm{num,\sigma_R}}$
and $T_{\mathrm{num},\beta}$ suggests that the real blow-up time may jump to different numerical
values as the nonlinear parameters $\beta, \sigma_R$ vary (the behavior for $\mu, \nu$ is similar),
although remaining constant within specific intervals of variation.
This effect may be related to  transition between different dynamics of the
initial pulse prior to collapse, as will be shown below.

In Fig.~\ref{fig5}, we present the results of a numerical study verifying 
that (\ref{ap4Icrit}) is an approximate critical
point separating global existence from collapse. Parameter values used for this study are
$\beta=0.02$, $\delta=0.01$, $\epsilon=1$, $\mu=\nu=0.01$, $\sigma_R=0.2$ and $L=50$.
The left upper panel illustrates the evolution of the initial density (\ref{pw1a1}) for $t\in[0,15]$, with $\gamma=0.1>\gamma^*\approx 0$, towards collapse at time $T_{\mathrm{num}}=27.85$ ($< T_{\mathrm{max}}=31$
---the analytical upper bound). We observe a nontrivial collapse mechanism, characterized (after a transient stage) by a self-similar evolution.
\begin{figure}
	\begin{center}
		\begin{tabular}{c}
			\includegraphics[scale=0.46]{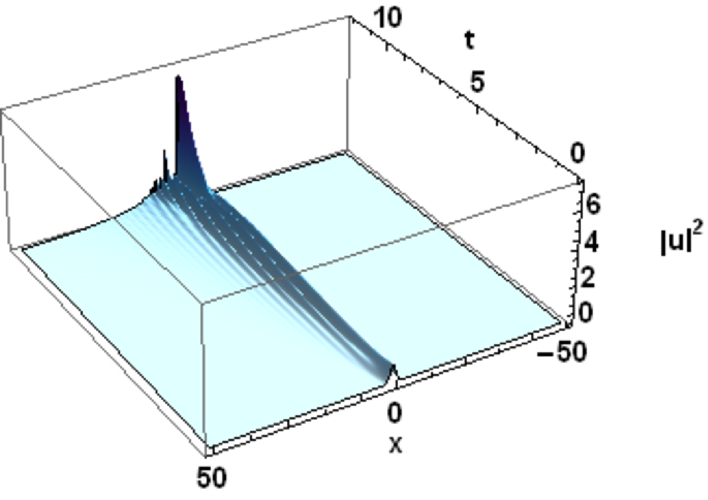}
			\includegraphics[scale=0.46]{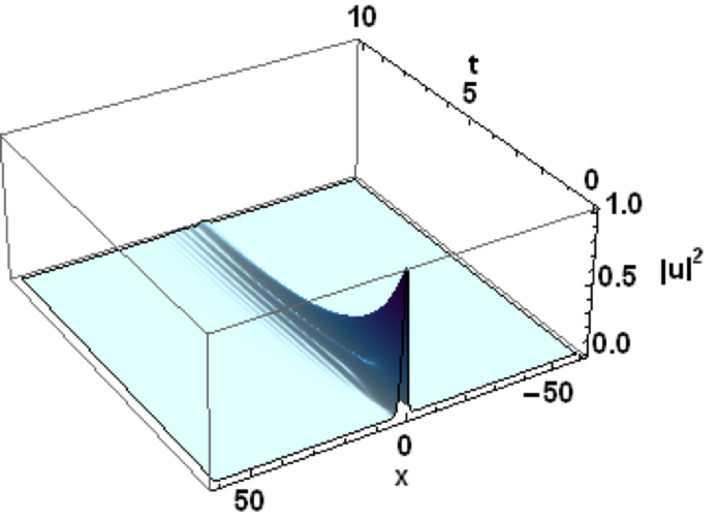}\\
			\hspace{0.5cm}\includegraphics[scale=0.55]{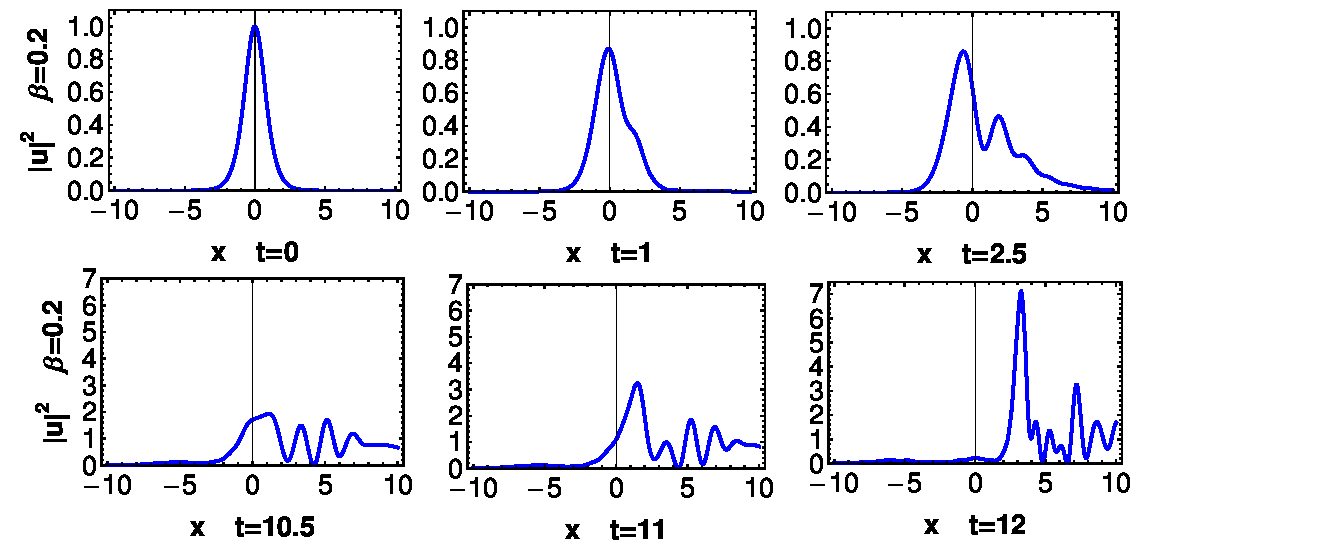}\\		
		\end{tabular}
		\caption{(Color Online) Dynamics for $\mathrm{sech}$-type initial data, when  $\beta=0.2$. The rest of parameters are fixed as in Fig. \ref{fig5}. Upper left panel: Evolution of the density $|u|^2$ towards collapse, when $\gamma=0.1>\gamma^*\approx0$. Upper right panel: Decay of the density $|u|^2$ when $\gamma=-0.1<\gamma^*\approx 0$.
			Bottom panel: Snapshots of the density $|u|^2$ towards collapse when $\gamma=0.1>\gamma^*\approx0$, corresponding to the density evolution shown in the upper left panel.}
		\label{fig6}
	\end{center}
\end{figure}
Initially,
the pulse amplitude is decreased but its width is increased (i.e., the
pulse tends to disperse); nevertheless,
as the gain mechanisms start playing a crucial role in the dynamics,
the effect on the amplitude is reversed, and both
width and amplitude grow self-similarly towards
a collapsing state. To highlight this behavior,
we show in the bottom panel the evolution of  $|u(0,t)|$ of the center of pulse for $t\in[0,20]$. The amplitude decays and after reaching a minimum, it starts growing. The insets are showing specific snapshots of the density $|u|^2$ corresponding to specific times and values of norm. The first one,
shows the initial condition at $t=0$, while the second shows the density profile at $t=4$, where the amplitude
has decreased and the width of the pulse has increased. The last one shows the density profile at $t=13$,
where both amplitude and width have increased, and the pulse evolves
towards collapse. On the other hand,
as shown in the upper right panel of Fig.~\ref{fig5}, when $\gamma=-0.1<\gamma^*$ the initial density decays
at an exponential rate. This decay still occurs in a self-similar manner, where the amplitude of the
pulse decays faster than the increase of its width.
Similar dynamics towards  collapse (or decay) have been observed for small variations of $\beta$,
as well as of the rest of the nonlinear parameters.

Larger variations of these parameters may lead to a different type of transient dynamics towards collapse or decay.
In Fig.~\ref{fig6}, we present the study for $\beta=0.2$ and the rest of parameters fixed as in
Fig.~\ref{fig5}. The 
top left panel shows the  evolution of the initial density (\ref{pw1a1}) for $t\in[0,12]$  towards collapse.
The numerical blow-up time is decreased further to $T_{\mathrm{num}}=20.89$, and the dynamics is very
different from the self-similar evolution of the previous case, due to the considerable dispersion effect.
This difference is highlighted in the bottom panel, showing snapshots of the density profiles at specific times, prior to collapse. The initial pulse is decomposed to a ``core pulse'' of decreased amplitude and a nonlinear wave train,
both traveling to the right at slow speed (upper row).  Progressively (bottom rows), and as the gain becomes more prominent, the amplitudes of the core and of the decomposed waves start increasing (note the change of scaling in the amplitude axis). While at a particular time instant, the amplitude of the wave-train is almost reaching that of the core, as shown in the snapshot
for $t=10.5$, eventually the core gains amplitude in comparison with the rest of the waves, as shown in the snapshot
for $t=11$. A sudden increase of the amplitude of the core is observed at $t=12$, leading to collapse.
For further variations of the parameter $\beta$, the numerical blow-up time is stabilized ---an effect illustrated
in the inset of the right panel of Fig.~\ref{fig4a}--- and the collapse dynamics were found to be similar.
The top right panel present the results of the numerical study for $\gamma=-0.1<\gamma^*$. The initial pulse is again decomposed to a wave train, but now the dynamics are reversed to decay instead of collapse.

\subsection{Non decaying (tanh$^2$-profiled) initial conditions.}
The third example of initial condition which we shall examine is the case of
tanh$^2$-profiled initial data,
\begin{eqnarray}
\label{tana1}
u_0(x)=\epsilon\, \mathrm{tanh}^2 x,
\end{eqnarray}
of amplitude $\epsilon>0$ 
(here, the discussion on the values of the derivatives on the boundaries should be recalled). 
Such initial data resemble a
density ($|u|^2$) profile reminiscent of the intensity dip of a dark soliton of the defocusing NLS,
on the top of the cw background, which is compliant with our
periodic boundary conditions. Note, that a simpler $\tanh$-type initial condition (real or complex
to model the black or gray solitons of the NLS) would violate our periodic boundary conditions
(and would require a different type of boundary conditions, such as
Neumann). While the observations here are still valid for the latter case,
we will report corresponding results in a future communication.
For the initial data (\ref{tana1}),
\begin{eqnarray*}
	M(0)=\frac{1}{2L}\int_{-L}^{L}\epsilon^2\mathrm{tanh}^4 xdx=\epsilon^2\left[\frac{3L-(3+\mathrm{tanh}^2L)\mathrm{tanh}L}{3L}\right].
\end{eqnarray*}
Thus, in the present case, Theorem \ref{T1a} and (\ref{eqUB})
assert that the analytical upper bound of the collapse time $T_{\mathrm{max}}[\gamma,\delta, M(0)]$ is
\begin{eqnarray}
\label{tana2}
T_{\mathrm{max}}[\gamma,\delta,\epsilon,L]=\frac{1}{2\gamma }\log \left[ 1+\frac{3\gamma L}{\delta\epsilon^2[(3L-(3+\mathrm{tanh}^2L)\mathrm{tanh}L]} \right].
\end{eqnarray}
Then, (\ref{eqUB0}) asserts that the analytical upper bound for the collapse time in the case $\gamma=0$ is
\begin{eqnarray}
\label{tana3}
T_{\mathrm{max}}[\delta,\epsilon,L]=\frac{3L}{2\delta \epsilon^2[(3L-(3+\mathrm{tanh}^2L)\mathrm{tanh}L]}.
\end{eqnarray}
Furthermore, Eq.~(\ref{CRIT}), implies that the critical value $\gamma^*$
in the present case [see (\ref{tana2})] is:
\begin{eqnarray}
\label{tana4}
\gamma^*=-\delta\epsilon^2\left[\frac{3L-(3+\mathrm{tanh}^2L)\mathrm{tanh}L}{3L}\right].
\end{eqnarray}

It is interesting to recover that for a sufficiently large
$L$, the initial value of the functional $M(t)$ at $t=0$ (\ref{tana1}), the analytical upper bounds for the collapse time (\ref{tana2}) and (\ref{tana3}),
and the critical value $\gamma^*$ given in (\ref{tana4}), approximate the relevant values for the example of the plane-wave initial conditions discussed in Sec.~3.1. Namely, for 
large $L$,
\begin{eqnarray}
\label{tana6}
&&M(0)\approx\epsilon^2,\;\;T_{\mathrm{max}}[\gamma,\delta,\epsilon,L]\approx \frac{1}{2\gamma }\log \left[ 1+\frac{\gamma}{\delta\epsilon^2} \right],\;\;\mbox{for $\gamma\neq 0$},\\
\label{tana7}
&&T_{\mathrm{max}}[\gamma,\delta,\epsilon,L]\approx\frac{1}{2\delta \epsilon^2},\;\;\mbox{for $\gamma\neq 0$},\;\;\mbox{and}\;\;
\gamma^*\approx -\delta\epsilon^2.
\end{eqnarray}
This result is of course not surprising given that the fact that
as $L$ increases the weight nature of the average in the definition of
$M(0)$ yields a progressively vanishing contribution from the localized
part of the relevant initial condition profile.
The approximations (\ref{tana6}) and (\ref{tana7}) may indicate that the numerical collapse times may not only respect, but also be much closer to their analytical upper bounds as in the case of the plane-wave initial data. Another feature, interesting by itself, which may support the above argument, is that the collapse may be manifested by the collapse
of the wave background for small values of the dispersion parameter $\beta$ and of the nonlinear parameters,
as discussed in Sec.~3.1. This fact should be also supported by the expectation that the dip of the density profile
should have a little effect to the averaged norm $M(t)$, and the main contribution should
come from the wave background. More precisely, the approximate equation on the evolution of $M(t)$
should be also valid in the case of the initial data (\ref{tana1}).

\begin{figure}[ht]
	\begin{center}
		\begin{tabular}{c}
			\hspace{2.5cm}\includegraphics[scale=0.55]{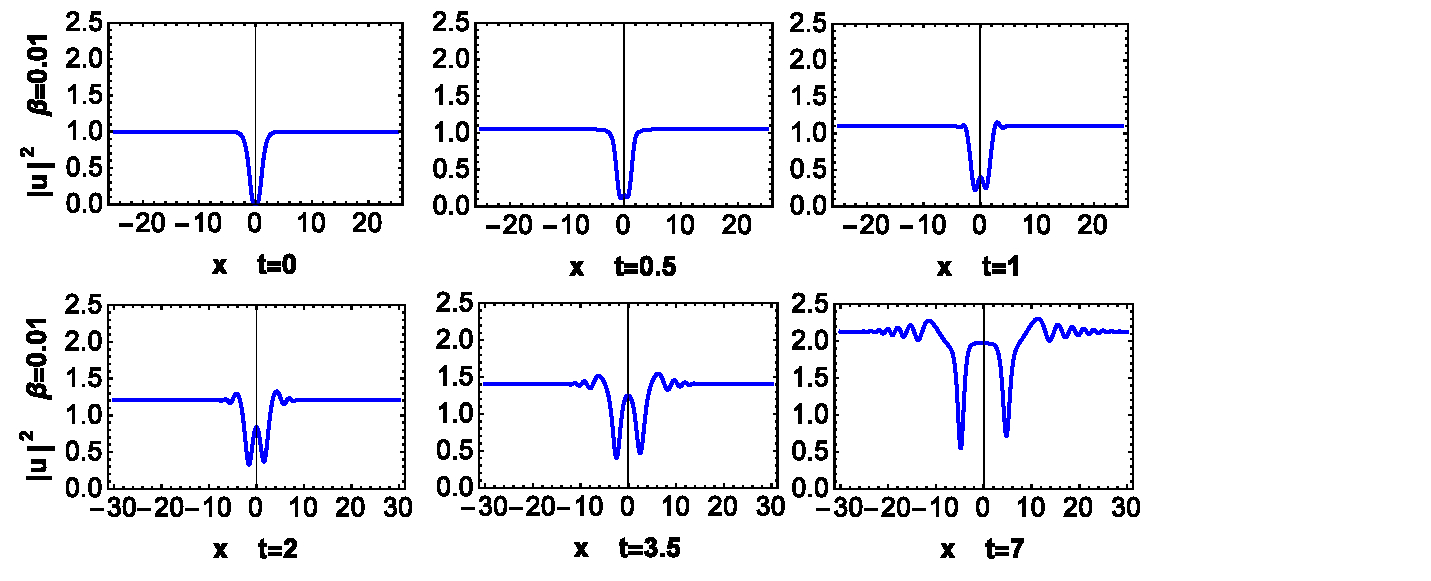}\\
			\includegraphics[scale=0.48]{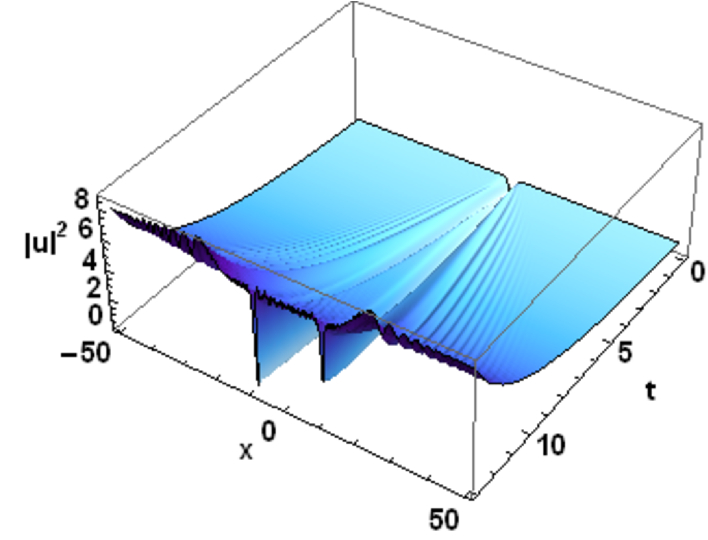}
			\includegraphics[scale=0.4]{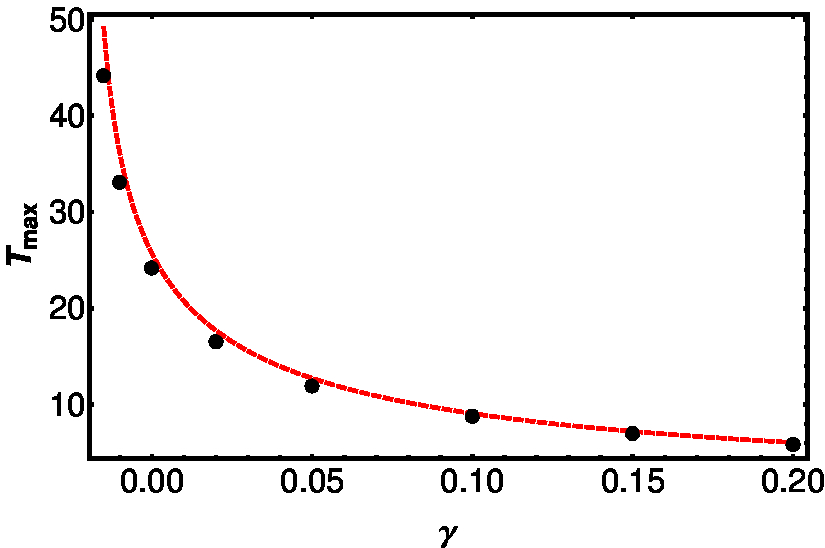}
		\end{tabular}
		\caption{(Color Online) Upper panel: Snapshots of the evolution of the density $|u|^2$ towards collapse for $\mathrm{tanh^2}$-type initial data, when $\delta=0.02$, $\epsilon=1$ and $\gamma=0.025>\gamma^*=-0.02$.  The rest of parameters are $L=50$, $\mu=0.01$, $\nu=0.01$, $\sigma=0.02$, $\beta=0.01$.  Bottom left panel: An upper view for the collapse of the density $|u|^2$ for $\mathrm{tanh^2}$-type initial data for the set of parameters of the upper panel.  Bottom right panel: Comparison of the analytical upper bound  (\ref{tana2}) (dashed (red) curve) against the numerical blow-up time (dots) as a function of $\gamma$. Rest of parameters are fixed as in the study of the upper panel.}
		\label{fig7}
	\end{center}
\end{figure}

In the top and bottom left panel of Fig.~\ref{fig7}, we present the numerical results on the evolution of the density of initial data (\ref{tana1}), for $\epsilon=1$, $\gamma=0.025$ and $\delta=0.02$.  The rest of parameters are $\mu=0.01$, $\nu=0.01$, $\sigma=0.02$, $\beta=0.01$, and the system is integrated for $L=50$.
For the given values of parameters, the critical value $\gamma\approx -0.02<0$, and since $\gamma>\gamma^*$, we are in the regime of finite-time collapse. The top panel shows snapshots of the density profiles at specific times, and the bottom left panel shows a (bottom) view of the density evolution, prior to collapse. The increase of the density is manifested by the increase of the height of the wave-background and accompanied by the nontrivial dynamics of the initial dip,
prior to collapse at numerical time $T_{\mathrm{num}}=15.9$,
while the analytical upper bound is $T_{\mathrm{max}}=16.5$.
As highlighted in the upper panel, the initial density dip appears to
separate into two traveling (at low speed) dark pulses of progressively
increasing depths, emitting radiation in the form of linear waves of increasing amplitude. In fact, the emerging patterns bearing an oscillatory
wavefront, together with a solitary wave are reminiscent of dispersive
shock waves (DSWs)~\cite{DSW}.
The generation of two dark soliton pulses can be understood by the fact that the initial condition
has a trivial phase profile (i.e., does not have the phase jump needed for a dark soliton); thus,
the initial dip separates into two dark solitons moving to opposite directions so that the phase of
one soliton is compensated
by the other soliton, with the total phase retaining its initial
asymptotic structure -- see, e.g., Ref.~\cite{Blow}. On the other hand,
the clearly observed asymmetry in the evolution profile is apparently
produced by parity breaking terms, such as the first and third spatial
derivatives in Eq.~(\ref{eq1}).
The bottom right panel of Fig. \ref{fig7} shows the comparison of the analytical upper bound (\ref{tana2}) (dashed (red) curve) against the numerical blow-up time (dots) as a function of $\gamma$, when the rest of parameters are fixed as in the study of the upper panel. We observe that the numerical blow-up  are
quite  close to their analytical upper bounds, similarly (for the
reasons indicated above) to the plane wave case.

\begin{figure}[h]
	\begin{center}
		\begin{tabular}{c}
			\includegraphics[scale=0.5]{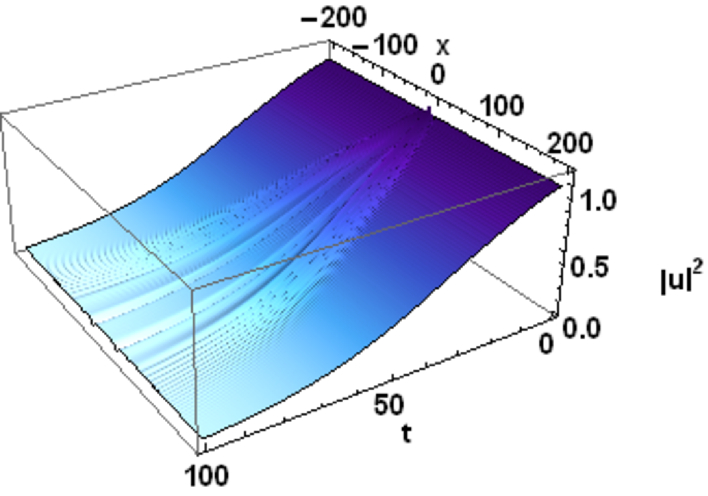}
			\includegraphics[scale=0.52]{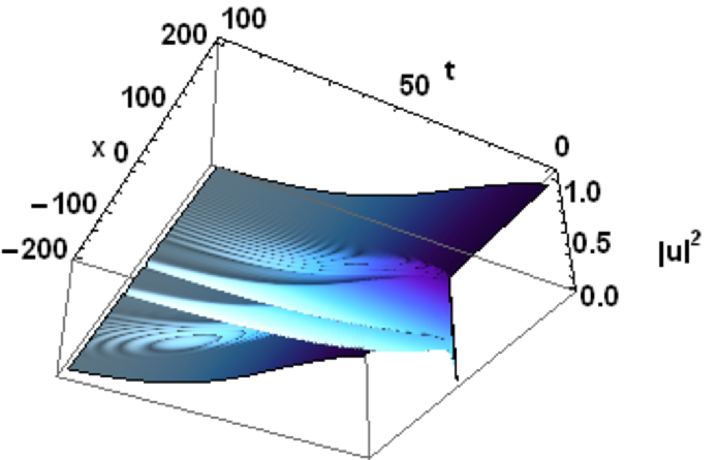}
		\end{tabular}
		\caption{(Color Online) Left panel: An upper view for the decay of the density $|u|^2$ for $\mathrm{tanh^2}$-type initial data, as found for  $\delta=0.02$, $\epsilon=1$, and  $\gamma=-0.025<\gamma^*=-0.02$. The rest of parameters are $L=200$, $\mu=0.01$, $\nu=0.01$, $\sigma=0.02$, $\beta=1$. Right panel: A bottom view for the decay of the density $|u|^2$ for $\mathrm{tanh^2}$-type initial data. Parameters as in the upper panel.  }
		\label{fig8}
	\end{center}
\end{figure}

In Fig.~\ref{fig8}, keeping the rest of parameters as above and integrating the system for $L=200$,
we present the results of the numerical study for slightly perturbing the parameter $\gamma$ from its critical value $\gamma^*=-0.02$, falling
within the decay regime, e.g, by considering $\gamma=-0.025$.
The left (right) panel shows a top (bottom) view of the evolution of the density manifested
by the decay of the height of the wave-background again accompanied by nontrivial dynamics of the initial dip,
prior to decay:
it separates again into two traveling (at low speed) dark pulses, this time of slowly decreasing depths.
The two decaying pulses are emitting radiation again, which also decays, following the decay of the wave-background. Once again, on both sides
the oscillatory waveforms coupled to the dark solitary waves are
somewhat reminiscent of DSW patterns.

\begin{figure}[h]
	\begin{center}
		\begin{tabular}{c}
			\hspace{2.5cm}\includegraphics[scale=0.55]{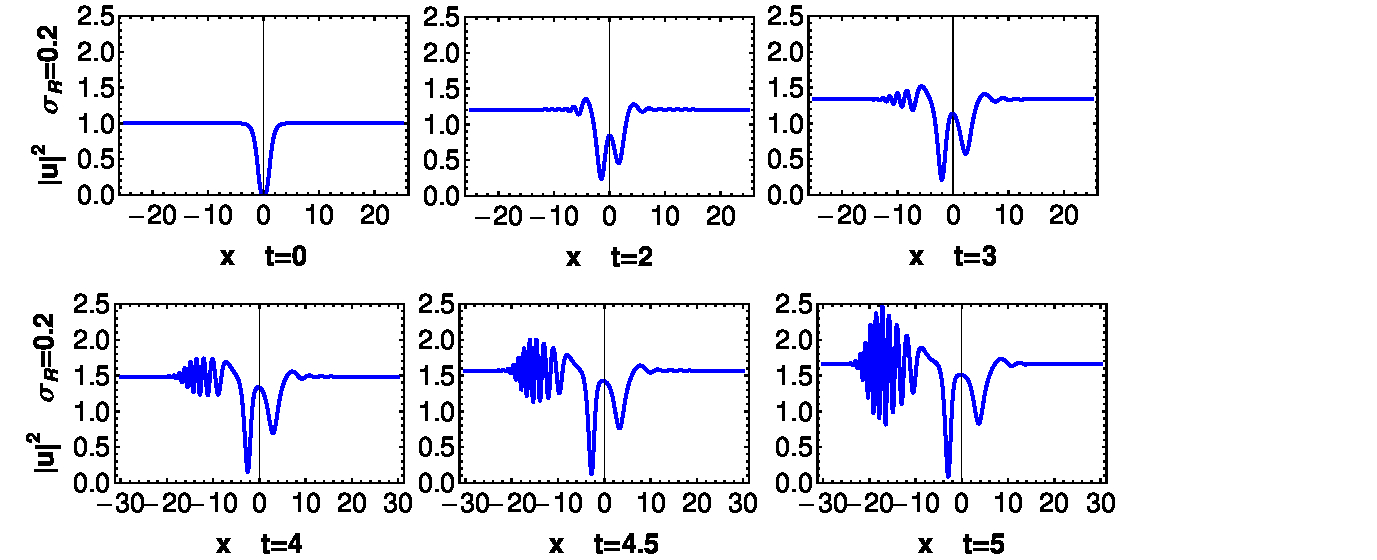}\\		
		\end{tabular}
		\caption{(Color Online) Snapshots of the evolution of the density $|u|^2$ towards collapse when $\gamma=0.025>\gamma^*=-0.02$, all the parameters fixed as in the study of Fig. \ref{fig7}, but for the increased value of $\sigma_R=0.2$.}
		\label{fig9}
	\end{center}
\end{figure}

For larger valued-parametric regimes of the higher-order effects, the transient dynamics prior to collapse or decay
were found to be potentially quite different. Concerning collapse, we have recovered the same
effect for the variation of the numerical blow-up time associated with the transition to a different type of
collapsing dynamics. In Fig.~\ref{fig9}, we present snapshots of the density evolution for $\sigma_R=0.2$ and the rest of the parameters fixed as in the case of Fig.~\ref{fig7} ---the case of collapse. While the initial dark pulse
is still separated into two dips as explained above, we also observe
stronger ``one-sided'' radiation to the left,
due to the third-order dispersion, which is increased due to the (relatively strong) Raman effect.
The two dips and the right-part of the background seem to have a similar
dynamics as in the case of Fig.~\ref{fig7}, however the collapse is manifested by the sudden increase of the amplitudes of the
oscillatory waveforms in the background
formed to the left of the dips.

\begin{figure}
	\begin{center}
		\begin{tabular}{c}
			\includegraphics[scale=1]{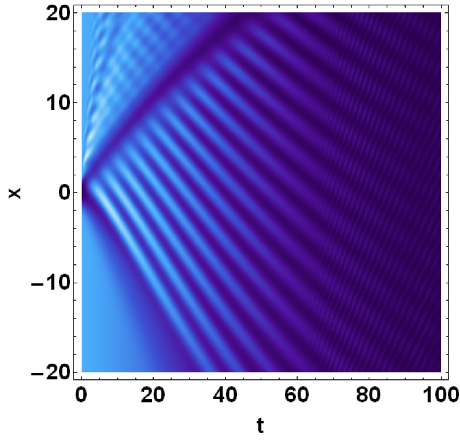}\vspace{-0.2cm}\\
		\hspace{0.7cm}\includegraphics[scale=0.45]{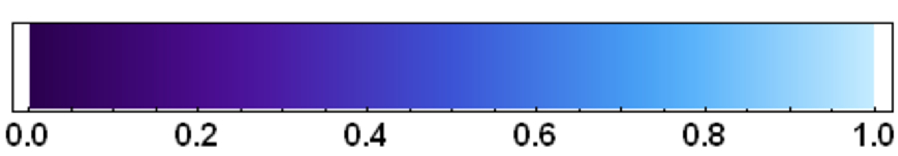}
		\end{tabular}
		\caption{(Color Online) Contour plot of the evolution of the density $|u|^2$ for $\gamma=-0.025<\gamma^*$ and $\beta=0.2$. The rest of
the parameters is fixed as in the  the example of Fig. \ref{fig8}.}
		\label{fig10}
	\end{center}
\end{figure}

The transient dynamics of our system
can be very rich. As an example, in Fig.~\ref{fig10} we
show a contour plot of the space-time evolution of the density $|u|^2$ in the decay regime
$\gamma=-0.025<\gamma^*=-0.02$ (the rest of parameters fixed as in the study of Fig.~\ref{fig8}), but for increased
dispersion effect, namely $\beta=0.2$. The initial dip is decomposed to an almost spatially periodic pattern,
traveling to the left. These periodic structures, in turn,
have different survival times prior to the their final decay.
A more detailed study of such phenomenology is warranted although
it is outside the scope of the present investigation.

\section{Conclusions}
In this work, analytical studies corroborated by numerical simulations, considered the collapse dynamics of the higher-order NLS equation (\ref{eq1}) supplemented by periodic boundary conditions. The model has important applications in describing the evolution of ultrashort pulses in nonlinear media (such as optical fibers, metamaterials, water waves),
where linear and nonlinear gain/losses and higher order conservative and dissipative effects are of importance. We have shown by means of simple analytical energy arguments and verified numerically, that the collapse/decay dynamics are chiefly dominated by the linear and nonlinear gain/loss  of strength $\gamma$ and $\delta$ respectively.  We identified a critical value $\gamma^*$ in the regime $\gamma<0$, $\delta>0$,  which is separating finite-time collapse from the decay of solutions. An important conclusion regarding this regime, is related to the fate of localized solutions such as pulse and dip-type waveforms \cite{DJFTheo}: referring to their true long-time dynamics, such states only
survive for relatively short times, as they are subject to either
catastrophic collapse or to eventual decay for generic parameter values.

The results of the numerical simulations for a variety of initial conditions
suggest a very good agreement between the analytical estimates for the collapse times and the numerical ones, as well on the accuracy of the critical value $\gamma^*$. This agreement was excellent for plane wave initial data, and
very good for waveforms bearing a density dip, while it was less good in the
case of initial data resembling bright solitary wave structures, for
reasons explained in the text.
Furthermore, the numerical simulations gave insight on the role of the strengths of the higher order effects on the transient dynamics prior to
collapse or decay, which were found to be essentially different.
They have also indicated that these strengths may seriously affect the
survival times of the solutions.

Our results pave the way for a better understanding of the dynamics of the considered higher-order 
NLS equation. 
An interesting direction is to discuss the collapse dynamics for the case of $s=-1$. 
It is also certainly relevant, to explore the associated dynamics in the stabilization regime $\gamma>0$, $\delta<0$ respectively (see \cite{PartII} for some recent results), depending on the 
higher-order effects and the type of initial data. Future investigations should also 
consider the possibility of collapse of higher-order norms, and the effect on this type of collapse of the higher-order nonlinear terms, as well as of higher-dimensional settings, 
which may further modify the dynamical regimes analyzed herein.

\section*{Acknowledgements}

P.G.K. gratefully acknowledges the support of
the US-NSF under the award DMS-1312856,
and of the ERC under FP7, Marie
Curie Actions, People, International Research Staff
Exchange Scheme (IRSES-605096).
P.G.K.'s work at Los Alamos is supported in part by the U.S. Department
of Energy. The work of D.J.F. was partially supported by the Special Account for
Research Grants of the University of Athens.

\appendix

\section{Derivation of Eq.~(\ref{OL1N})}

Here, for the sake of completeness, we provide details for the derivation of Eq.~(\ref{OL1N}) and also
explain the choice of the functional $M(t)$ [cf. Eq.~(\ref{eq4a})].

First we note that the second term on the right-hand side of Eq.~(\ref{OL1}),
after substituting $u_t$ by the right-hand side of Eq.~(\ref{eq1}) becomes:
	\begin{eqnarray}
	\label{OL2}
	\frac{e^{-2\gamma t}}{L}\mathrm{Re}\int_{-L}^{L}u_t\overline{u}dx=\frac{e^{-2\gamma t}}{L}\sum_{k=1}^7 J_k,
	\end{eqnarray}
where
	\begin{eqnarray*}
		&&J_1=-\mathrm{Re}\left\{\frac{\mathrm{i}}{2}\int_{-L}^{L}u_{xx}\overline{u}dx\right\},\,J_2=\mathrm{Re}\left\{\mathrm{i}\int_{-L}^{L}|u|^2u\overline{u}dx\right\},\nonumber\\
		&&J_3=\mathrm{Re}\left\{\gamma\int_{-L}^{L}u\overline{u}dx\right\},\qquad
		J_4=\mathrm{Re}\left\{\delta\int_{-L}^{L}|u|^2u\overline{u}dx \right\},\\
		&&J_5=\mathrm{Re}\left\{\beta\int_{-L}^{L}u_{xxx}\overline{u}dx\right\},\;\;J_6=\mathrm{Re}\left\{\mu\int_{-L}^{L}(|u|^2u)_x\overline{u}dx\right\},\\
		&&J_7=\mathrm{Re}\left\{\int_{-L}^{L}(\nu-\mathrm{i}\sigma_R)(|u|^2)_xu\overline{u}dx\right\}.
	\end{eqnarray*}
	Integration by parts and use of
periodicity, implies that $J_5=-J_5$, hence $J_5=0$. By the same token, we have that $\int_{-L}^{L}(|u|^2)_x|u|^2dx=0$, hence,
	\begin{eqnarray*}
		J_6&=&\mu\mathrm{Re}\int_{-L}^{L}(|u|^2)_xu\overline{u}dx
		+\mu\mathrm{Re}\int_{-L}^{L}|u|^2u_x\overline{u}dx\\
		&=&\mu\mathrm{Re}\int_{-L}^{L}(|u|^2)_x|u|^2dx+\mu\int_{-L}^{L}|u|^2\mathrm{Re}(u_x\overline{u})dx\\
		&=&\mu\mathrm{Re}\int_{-L}^{L}(|u|^2)_x|u|^2dx+\frac{\mu}{2}\mathrm{Re}\int_{-L}^{L}(|u|^2)_x|u|^2dx=0,\\
		J_7&=&\mathrm{Re}\left\{(\nu-\mathrm{i}\sigma_R)\int_{-L}^{L}(|u|^2)_x|u|^2dx\right\}=0.
	\end{eqnarray*}
	Similarly, we have that $J_1=\mathrm{Re}\left\{\frac{\mathrm{i}}{2}\int_{-L}^{L}|u_x|^2dx\right\}=0$, and $J_2=\mathrm{Re}\left\{\mathrm{i}\int_{-L}^{L}|u|^4dx\right\}=0$, as well. Thus, we recover that the only contribution to (\ref{OL2}) comes from the terms $J_3=\gamma\int_{-L}^{L}|u|^2dx$, and $J_4=\delta\int_{-L}^{L}|u|^4dx$.
	Inserting all the above relations to (\ref{OL2}), we observe that the latter is actually
	\begin{eqnarray}
	\label{OL2N}
	\frac{e^{-2 \gamma t}}{L}\mathrm{Re}\int_{-L}^{L}u_t\overline{u}dx=
	\gamma\frac{e^{-2\gamma t}}{L}\int_{-L}^{L}|u|^2dx+\delta\frac{e^{-2\gamma t}}{L}\int_{-L}^{L}|u|^4dx,
	\end{eqnarray}
and consequently, Eq.~(\ref{OL1}) results in Eq.~(\ref{OL1N}).

Finally, we remark that following computations similar to the above, but for the
(squared) $L^2$-norm functional, we find that:
%
\begin{equation}
\frac{d}{dt}\int_{-L}^{L} |u|^2 dx = 2\gamma \int_{-L}^{L} |u|^2 dx +2\delta \int_{-L}^{L} |u|^4 dx.
\label{cl}
\end{equation}
Thus, it is evident that the factor $\exp(-2\gamma t)$ is the integrating factor for the linear part of the
above equation. For $\gamma=\delta=0$, Eq.~(\ref{cl}) is nothing but the conservation of the $L^2$ norm in
the standard NLS model.

\section*{References}

\bibliographystyle{elsarticle-num}

\begin{thebibliography}{22}
\providecommand{\natexlab}[1]{#1}
\expandafter\ifx\csname urlstyle\endcsname\relax
  \providecommand{\doi}[1]{doi:\discretionary{}{}{}#1}\else
  \providecommand{\doi}{doi:\discretionary{}{}{}\begingroup
  \urlstyle{rm}\Url}\fi

\bibitem[{Hasegawa \& Kodama(1996)}]{KodHas87} A.  Hasegawa and Y. Kodama, {\em Solitons in optical communications},
Oxford Univeristy Press, 1996.
\bibitem[{Agrawal(2003)}]{Agra1} G. P. Agrawal, {\em Nonlinear Fiber Optics}, Academic Press, 2012.
\bibitem[{Kivshar \& Agrawal(2003)}]{Agra2} Yu. S. Kivshar and G. P. Agrawal, {\em Optical Solitons: From Fibers to Photonic Crystals}, Academic Press, 2003.

\bibitem[{Leblond \& Mihlache(2013)}]{Mih1} H. Leblond and D. Mihalache, Phys. Rep. \textbf{523} (2013), 61.

\bibitem[{Frantzeskakis \emph{et~al.}(2014)}]{Mih2} D. J. Frantzeskakis, H. Leblond and D. Mihalache,
Rom. Journ. Phys. {\bf 59} (2014), 767. 

\bibitem[{Scalora \emph{et~al.}(2005)}]{p31} M. Scalora, M. S. Syrchin, N. Akozbek, E. Y. Poliakov, G. D’Aguanno, N. Mattiucci, M. J. Bloemer, and A. M.
Zheltikov, Phys. Rev. Lett. \textbf{95} (2005), 013902.
\bibitem[{Wen \emph{et~al.}(2007)}]{p32} S. Wen, Y. Xiang, X. Dai, Z. Tang, W. Su, and D. Fan, Phys. Rev. A \textbf{75} (2007), 033815.
\bibitem[{Tsitsas \emph{et~al.}(2009)}]{p33} N. L. Tsitsas, N. Rompotis, I. Kourakis, P. G. Kevrekidis, and D. J. Frantzeskakis, Phys. Rev. E \textbf{79} (2009), 037601.

\bibitem[{Johnson (1977)}]{johnson} R. S. Johnson, Proc. R. Soc. Lond. A {\bf 357} (1977), 131.

\bibitem[{Sedletsky (2003)}]{sedletsky} Yu. V. Sedletsky, J. Exp. Theor. Phys. {\bf 97} (2003), 180.

\bibitem[{Slunyaev (2005)}]{slun} A. V. Slunyaev, J. Exp. Theor. Phys. {\bf 101} (2005), 926.

\bibitem[{Ikeda \emph{et~al}(1995)}]{p35} H. Ikeda, M. Matsumoto, and A. Hasegawa, Opt. Lett. \textbf{20} (1995), 1113.

\bibitem[{Suzuki(1971)}]{suz71} T. Tsuzuki, J. Low Temp. Phys. \textbf{4} (1971), 441.
\bibitem[{Hasegawa \& Tappert(1973)}]{Has73} A. Hasegawa and F. Tappert, Appl. Phys. Lett. \textbf{23} (1973), 171.
\bibitem[{Kivshar \& Luther Davies(1998)}]{Kiv98} Yu. S. Kivshar and B. Luther-Davies, Phys. Rep. \textbf{298} (1998), 81.

\bibitem[{Kevrekidis \emph{et~al.}(2008)}]{KFG1} P. G. Kevrekidis, D. J. Frantzeskakis, and R. Carretero- Gonz\'{a}lez, {\em Emergent Nonlinear Phenomena in Bose-Einstein Condensates. Theory and Experiment}, Springer-Verlag, 2008.

\bibitem[{Carretero- Gonz\'{a}lez \emph{et~al.}(2008)}]{KFG2} R. Carretero- Gonz\'{a}lez, D. J. Frantzeskakis,
P. G. Kevrekidis, Nonlinearity {\bf 21} (2008) R139.

\bibitem[{Frantzeskakis(2010)}]{DJF} D. J. Frantzeskakis. J. Phys. A \textbf{43} (2010), 213001.

\bibitem[{Horikis \& Frantzeskakis(2013)}]{DJFTheo} T. P. Horikis and D. J. Frantzeskakis, Opt. Lett. \textbf{38} (2013), 5098.
%
\bibitem[{Ablowitz \emph{et~al.}(2011)}]{MJA} M. J. Ablowitz, S. D. Nixon, T. P. Horikis, and D. J. Frantzeskakis,
Proc. R. Soc. A (2011) {\bf 467}, 2597.

\bibitem[{Achilleos \emph{et~al.}(2015) }]{PartII} V. Achilleos, A. R. Bishop, S. Diamantidis, D. J. Frantzeskakis, T. P. Horikis,
N. I. Karachalios, and P. G. Kevrekidis,
arXiv:1509.03828 (2015).

\bibitem[{Kato(1975)}]{Kato0} T. Kato, {\em Quasilinear equations of evolution with applications to partial differential equations}, Lecture Notes in Mathematics \textbf{448}, pp. 25--70, Springer-Verlag, New York, 1975.

\bibitem[{Kato(1985)}]{Kato2}
T. Kato, {\em Abstract differential equations and nonlinear mixed problems}, Lezione Fermiane Pisa, 1985.

\bibitem[{Kato \& Lai(1984)}]{Katoat} T. Kato and C. Y. Lai, J. Funct. Anal. \textbf{56} (1984), 15. 

\bibitem[{Kato(1989)}]{Kato3} T. Kato, {\em Nonlinear Schr\"{o}dinger equations.
Schr\"{o}dinger operators}, S\o nderborg, 1988, 218--263, Lecture Notes in Phys.~\textbf{345}, Springer, Berlin, 1989.

\bibitem[{Cazenave(2003)}]{Caz03} T. Cazenave, {\em Semilinear Schr{\"o}dinger equations},
Courant Lecture Notes 10, Amer. Math. Soc., 2003.

\bibitem[{Ball(1977)}]{Ball77} J. M. Ball, Quart. J. Math. Oxford \textbf{28} (1977), 473.

\bibitem[{Sulem \& Sulem(1999)}]{Sulem99} C. Sulem and P. L. Sulem,
{\em The nonlinear Schr{\"o}dinger equation. Self-Focusing and wave collapse}, Springer-Verlag, 1999.

\bibitem[{Ozawa \& Yamazaki(2003)}]{OY03} T. Ozawa and Y. Yamazaki, Nonlinearity \textbf{16} (2003), 2029.

\bibitem[{Taylor(1996)}]{TaylorII} M. Taylor, {\em Partial Differential Equations III},
Applied Mathematical Sciences 117, Springer-Verlag, New York, 1996.

\bibitem[{Karachalios \emph{et~al.}(2007)}]{NET07} N. I. Karachalios, H. Nistazakis and A. Yannacopoulos,
Discrete Cont. Dyn. Syst. A \textbf{19} (2007), 711.

\bibitem[{Achilleos \emph{et~al.}(2013) }]{V13} V. Achilleos, A. {\'A}lvarez, J. Cuevas,
D. J. Frantzeskakis, N. I. Karachalios, P. G. Kevrekidis, and B. S{\'a}nchez-Rey, Physica D \textbf{244} (2013), 1.

\bibitem[{Hoefer and Ablowitz (2009)}]{DSW} M. Hoefer and M. Ablowitz,
Scholarpedia {\bf 4} (2009) 5562.

\bibitem[{Blow and Doran (1985)}]{Blow} K. J. Blow and N. J. Doran,
Phys.\ Lett.\ A, {\bf 107} (1985) 55. 

\end{thebibliography}

\label{lastpage}

\end{document}